\documentclass[12pt]{iopart}

\usepackage{lmodern}
\usepackage[T1]{fontenc}
\usepackage{amsmath, amsfonts, amsthm, amssymb, amstext, bm, bbm, graphics, fancyhdr, wasysym}

\DeclareSymbolFont{bbold}{U}{bbold}{m}{n}
\DeclareSymbolFontAlphabet{\mathbbold}{bbold}
 
 \DeclareMathAlphabet{\mathsfit}{T1}{\sfdefault}{\mddefault}{\sldefault}
\SetMathAlphabet{\mathsfit}{bold}{T1}{\sfdefault}{\bfdefault}{\sldefault}

\usepackage[usenames,dvipsnames]{xcolor}
\usepackage[hypcap]{caption}
\usepackage{subcaption, float, comment}
\usepackage{soul}

\usepackage[ hmarginratio={1:1}, vmarginratio={1:1}, textwidth=450pt, tmargin=1.0in, heightrounded ]{geometry}
\usepackage[colorlinks=true,linkcolor=blue,citecolor=red,urlcolor=blue,anchorcolor = blue,hypertexnames=true]{hyperref}

\usepackage[pagewise]{lineno}
\renewcommand{\theequation}{\thesection.\arabic{equation}}
 
\usepackage{soul}

\usepackage{titlesec}
\newcommand*{\justifyheading}{\raggedright}
\titleformat {\title}
{\normalfont\large\bfseries\justifyheading}{\thesection}{1em}{}
\titleformat {\section}
{\normalfont \large \bfseries \justifyheading}{\thesection}{1em}{}
\titleformat {\subsection}
{\normalfont \large \bfseries \justifyheading}{\thesubsection}{1em}{}
\titleformat {\subsubsection}
{\normalfont \bfseries \justifyheading}{\thesubsubsection}{1em}{}

\renewcommand\thesection{\arabic{section}}
\renewcommand\thesubsection{\thesection.\arabic{subsection}}
\renewcommand\thesubsubsection{\thesubsection\arabic{subsubsection}}

\newcommand{\gb}{$\mathbbm{g}$ }

\usepackage{tikz}
\usetikzlibrary{shapes.misc}
\usetikzlibrary{arrows}
\usetikzlibrary{arrows.meta}
\tikzset{
	tipA/.tip={Triangle[angle=45:10pt]}	
}

\tikzset{cross/.style={cross out, draw=black, minimum size=2*(#1-\pgflinewidth), inner sep=0pt, outer sep=0pt},
	cross/.default={1pt}}

\begin{document}

\title{Fixed Points and Universality Classes in Coupled Kardar-Parisi-Zhang Equations}
\author{Dipankar Roy}
\address{Université Côte d'Azur, CNRS, LJAD, 06108 Nice, France}
\author{Abhishek Dhar}
\address{International Centre for Theoretical Sciences, Tata Institute of Fundamental Research, Bangalore 560089, India}
\author{Manas Kulkarni}
\address{International Centre for Theoretical Sciences, Tata Institute of Fundamental Research, Bangalore 560089, India}
\author{Herbert Spohn}
\address{Zentrum Mathematik and Physik Department, Technische Universität München, Garching 85748, Germany}\vspace{80pt}
\textbf{Abstract}.     We study coupled KPZ equations with three control parameters $X,Y,T$. These equations are used in the context 
of stretched polymers in a random medium,  for the spacetime spin-spin correlator of the isotropic quantum Heisenberg chain, and for exciton-polariton condensates.
 In an earlier article we investigated merely
the diagonal $X=Y$, $T=1$. Then the stationary measure is delta-correlated Gaussian and  the dynamical exponent is obtained numerically to be close to $z = \tfrac{3}{2}$. We observed that the scaling functions of the dynamic correlator  change smoothly when varying $X$.
In this contribution, the analysis is extended to the whole $X$-$Y$-$T$ plane. Solutions are stable only if $XY \geq 0$. Based on numerical simulations, the static correlator still has rapid decay. We argue that the parameter space is foliated into distinct universality classes. They are labeled by $X$ and consist of half-planes parallel to the $Y$-$T$ plane containing the point $(X,X,1)$. 

\newpage


\section{Introduction}
\label{sec1}
\setcounter{equation}{0}
The KPZ equation was introduced by Kardar, Parisi, and Zhang~\cite{1986-kardar--zhang} as a model for growing surfaces. The equation governs the motion of a height
function, $h(x,t)$, where $x\in \mathbb{R}$ is the substrate coordinate and $t \geq 0$ time. In dimensionless  units the KPZ equation is written as   
\begin{equation}
    \label{1.1}
    \partial_t h(x,t) =  \lambda(\partial_x h(x,t))^2 +  \tfrac{1}{2}\partial_x ^2h(x,t) + \xi(x,t),
\end{equation}
where $\xi(x,t)$ is standard spacetime Gaussian white noise. The mean zero time-stationary measure is known to be standard spatial white noise,
which immediately implies the wandering exponent $\chi = \tfrac{1}{2}$. From the sum rule $\chi + z = 2$ \cite{forster,Krug1977,medina,2015-quastel-spohn}
one infers
the dynamical exponent $z = \tfrac{3}{2}$. On such scale there are universal distribution functions. For example, for flat initial conditions the height at the origin  
has fluctuations as
\begin{equation}
    \label{1.2}
h(0,t)  - \tfrac{1}{3}\lambda^3t \simeq \mathrm{sgn}(\lambda)\big(|\lambda|t/4\big)^{1/3} \xi_\mathrm{TW}    
\end{equation}
for large $t$. The random variable $\xi_\mathrm{TW}$ is distributed according to GOE Tracy-Widom~\cite{ps2002,2018-takeuchi}. For other 
models in the KPZ universality class the coefficient $\lambda$ will  be modified, but $\xi_\mathrm{TW}$ remains unaltered. Up to scale factors, the KPZ equation has a \textit{single} universality class.

Our interest is the same equation with several components. Using the Einstein summation convention and maintaining for the moment 
all model parameters, the generalization reads  
\begin{equation}
    \label{1.3}
    \partial_t h_\alpha =    G^\alpha_{\beta\gamma} (\partial_x h_\beta )(\partial_x h_\gamma) + \tfrac{1}{2} D_{\alpha\beta}\partial_x^2 h_\beta + B_{\alpha\beta}\xi_\beta,
\end{equation}
$\alpha,\beta,\gamma = 1,...,n$. 
Here the $n\times n$ matrices $G^\alpha$, $D$, and $B$ are given constants and the noise components $\xi_\alpha$ are independent.
In this note we will consider only $n = 2$. Even then there are still many model parameters. A physically natural simplification arises from studies 
in the context of nonlinear optics, which have 
the goal to experimentally confirm  KPZ physics. One favoured system are two coupled exciton-polariton condensates.
Their effective description through coupled KPZ equations is extensively covered in the recent article \cite{weinberger2024}. Since the condensates are physically indistinguishable, Eq. \eqref{1.3}  has to be invariant under the interchange 
$h_1\leftrightarrow  h_2$. Using this symmetry, it can be shown~\cite{weinberger2024} that the $h_\alpha$-fields, rotated by $\pi/4$ and appropriately rescaled, are governed by
\begin{equation}\label{1.4}
    \begin{aligned}
	\partial_t h_1 &=    2 X (\partial_x h_1)(\partial_x h_2) + \tfrac{1}{2}T\partial_x^2 h_1 +\sqrt{T}\xi_1  , \\
	\partial_t h_2 &=  Y  (\partial_xh_1)^2 + (\partial_xh_2)^2 + \tfrac{1}{2} \partial_x^2 h_2+ \xi_2 ,
    \end{aligned}
\end{equation}  
 where $X,Y$ are real parameters and $T>0$. Details are provided in \ref{A}.  
 These coupled KPZ equations first appeared in a study of Erta\c{s} and Kardar~\cite{1993-ertas-kardar},  who investigated the dynamics of stretched polymers immersed in a random medium. More recently the same equations reappeared in the study of 
sliding particles on a fluctuating landscape~\cite{2017-chakraborty--barma,2019-chakraborty--barma}. In the context of 
 quantum spin chains  of great interest is the equilibrium spacetime correlator of the isotropic Heisenberg chain~\cite{2023-nardis--vasseur}. In simulations, also in experiments, it was observed that  this correlator has the dynamical exponent $z = \tfrac{3}{2}$~\cite{2021-scheie--tennant}. Moreover the shape turned out to be in good agreement with the stationary scaling function of Eq. \eqref{1.1}.
 To explain such  findings a pair of coupled KPZ equations has been derived in~\cite{2023-nardis--vasseur}, which equals \eqref{1.4} upon setting  $T=1$.

In \eqref{1.1}, \eqref{1.3}, and \eqref{1.4} we have omitted a term linear in $\partial_x h_\alpha$. For the KPZ equation ($n=1$) this term is removed by a Galilei transformation. For $n \geq 2$ 
 the linear term would be of the form $F_{\alpha\beta}\partial_x h_\beta$. In general, the matrix $F$ is nondegenerate  and the linear term can no longer be transformed away. This results in $n$ modes separating ballistically in time. 
 Such case of nondegenerate eigenvalues has been studied in a variety of models \cite{SR1,SR2,2011-grisi-schutz,2013-ferrari--spohn,2013-mendl-spohn,2014-das--spohn,Dolai2024}. As worked out in great detail \cite{2014-spohn}, the universal features can then be reduced to the single component case. In accordance with the applications mentioned already, we set $F=0$ and require zero initial slope. This implies that all modes have zero velocity and thus interact strongly.
 
 For the KPZ equation the mean zero time-stationary measure can be computed explicitly. Switching to two components,  the mean zero time-stationary measure is not known. 
 Only under the condition $Y = X$ the $G^\alpha$ matrices are cyclic~\cite{2014-spohn,RDK24}, implying that the steady state consists of two independent copies of  spatial Gaussian white noise of strength $1$.
  To stress, the dynamics could be unstable and no such steady state would exist.  However, \textit{if} a mean zero time-stationary measure does exist and has rapid decay of correlations, then 
 the analogue of the wandering exponent still takes the value $\chi = \tfrac{1}{2}$. 
 To find out about the dynamical exponent, one observes that in the scaling regime
 diffusion and noise govern the microscopic scale, order of the correlation length, while the nonlinear terms govern the widely separated mesoscopic scale. Such reasoning does not  depend on the number of components. Thus, provided the steady state has short-ranged correlations, one would anticipate that Eq. \eqref{1.4} shows also the dynamical exponent $z = \tfrac{3}{2}$, see Section \ref{sec2.1} for a more complete discussion.
 
 Given the scaling exponent, finer details of universality are encoded in scaling functions. Of course, these depend on the particular observable. For $n=1$, there is a single universality class with an example provided in Eq. \eqref{1.2}, which refers to height fluctuations in case of flat initial 
 conditions. 
 For $n =2$, we focus on the 
 spacetime dynamic correlator of the two components of the slope field in the steady state. Structurally the equations become richer,
since one has to take into account linear combinations of the fields. Hence  the matrix of scaling functions is now considered to belong to the same universality class, if equal up to dilations and rotations. In principle, the scaling functions depend on all control parameters, $X,Y,T$. 
 Within the set of parameters satisfying $\chi = \tfrac{1}{2}$, the central issue is to understand how such scalings depend on model parameters.  To clarify this issue will be the main task of our contribution. 

\begin{figure}[!t]
\centering
    \includegraphics[width=7.5cm]{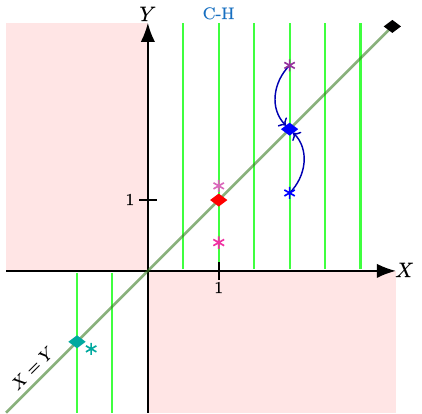}
    \caption{ Setting $T=1$, we display the anticipated phase diagram in the $XY$ plane.  The stable regime is $XY >0$, while the diagonal, $\{X=Y\}$, is a line of fixed points with distinct scaling functions.
    The green half-lines constitute the universality classes. 
    As exemplified in blue, starting with parameters on a particular green line the large time scaling function will be equal to the one of the corresponding fixed point, up to nonuniversal coefficients. The colored lozenges and asterisks indicate parameter points at which direct numerical simulations of Eqs. \eqref{1.4}  are reported, see Section 3. The black lozenge refers to the  limit along the diagonal, see Sections \ref{sec2.5} and \ref{sec2.8}. The half-line $\{X=1\}$ allows for a Cole-Hopf transformation, see Sections \ref{sec2.3}.}
\label{fig1}
\end{figure}
  
 As a guide to the reader, for $T = 1$, the anticipated phase diagram  is shown in Figure 1.  The diagonal $\{X=Y\}$ is a line of fixed points. 
 Along this line, scaling functions are smoothly varying in $X$. For general bare parameters $(X,Y)$, in the long time 
 limit the scaling functions converge to the ones of the diagonal point $(X,X)$. In Figure \ref{fig1}, this feature is indicated by the blue lines with arrow pointing towards the fixed point. We argue that the universality classes consist of the half-lines parallel to the $Y$-axis through parameter point $(X,X)$.  Distinct half-lines are separate universality classes. For general $X,Y,T$, the universality classes are argued to be  
 labeled by $X$ and to consist of half-planes parallel to the $Y$-$T$ plane containing the point $(X,X,1)$.
 
In an earlier contribution \cite{RDK24}, we studied universality  along the diagonal $\{X=Y\}$ with $T=1$ and confirmed that the scaling functions indeed change smoothly in their dependence on $X$. This
work expands the analysis to the full $ XY >0$ part of the plane and to general $T>0$. Our text is divided  into a theoretical and numerical part. In Section 2, we 
discuss stability, time-stationary
measures, Cole-Hopf transformation, the link to directed polymers, dynamical correlator, cyclicity, and the full 
phase diagram. We have included a subsection on the boundary points $X=0,Y>0 $ and $Y=0,X>0 $. Interestingly enough, their behavior foreshadows the transition to an unstable regime defined by $XY<0$.
In Section 3 we display and comment upon direct numerical simulations at the parameter points  marked in Figure 1 and beyond.
Finally, in the discussion section we compare our findings with the results based on the one-loop RG analysis in \cite{1993-ertas-kardar}, \cite{weinberger2024}. The heavily used rescalings are detailed in \ref{A}.

\section{Two coupled stochastic Burgers equations}
\label{sec2}
\setcounter{equation}{0} 
For our purposes it will be convenient to switch from height functions to slopes defined as $\phi_\alpha = \partial_xh_\alpha$.  Then the two component system under study is 
 \begin{equation}\label{2.1}
    \begin{aligned}
	\partial_t \phi_1 &= \partial_x \big(  2 X  \phi_1 \phi_2 + \tfrac{1}{2}T \partial_x \phi_1+ \sqrt{T}\xi_1 \big), \\
	\partial_t \phi_2 &= \partial_x \big( Y  \phi_1^2 + \phi_2^2 +  \tfrac{1}{2}\partial_x \phi_2  +\xi_2 \big),
    \end{aligned}
\end{equation}  
 where $X,Y$ are arbitrary real parameters and $T>0$. The noise $\xi_\alpha(x,t)$ is standard spacetime
 Gaussian white noise with covariance
  \begin{equation}
  \label{2.2}
  \langle \xi_\alpha(x,t) \rangle =0, \quad  \langle \xi_\alpha(x,t) \xi_\beta(x',t')\rangle = \delta_{\alpha\beta}\delta(x-x')\delta(t-t'). 
\end{equation}
 These are two stochastic conservation laws with quadratic nonlinearity, hence of stochastic Burgers type.

\subsection{Rescaling and $(1:2:3)$-scaling}
\label{sec2.1}
In Eq. \eqref{1.3} every term has a distinct strength parameter and  formulas tend to be
unwieldy. Thus it is more convenient to switch to dimensionless units. For this purpose
one chooses $\ell$ as unit of space, $\tau$ as unit of time, and the two field amplitudes as 
$h_\alpha(x,t) = a_\alpha \tilde{h}_\alpha (x/\ell,t/\tau)$.
 Then the form of the equation is not changed, but $a_\alpha,\ell,\tau$ can be used to reduce the number  of free parameters. For $n=2$ and symmetry 
$h_1\leftrightarrow  h_2$, the result is stated in 
Eqs. \eqref{1.4}, equivalently in Eqs. \eqref{2.1} and \eqref{2.2}, which have 3 dimensionless parameters. The respective computation is explained in \ref{A}.

A distinct argument concerns the separation between microscopic and mesoscopic scales. For such purpose we introduce  the dimensionless parameter $\varepsilon > 0$ and investigate the limit
$\varepsilon \to 0$. Conventionally used is the  scaled height function \cite{2015-quastel-spohn}, which is given by 
\begin{equation}
\label{2.2a}
h_{\varepsilon,\alpha}(x,t) = \varepsilon^{\chi} h_\alpha(\varepsilon^{-1} x, \varepsilon^{-z}t).
\end{equation}
Here $x,t$ are of order 1, $\chi$ is the wandering exponent, and $z$ the dynamical exponent. Using the results from \ref{A}, one arrives at 
\begin{equation}\label{2.3a}
    \begin{aligned}
	\partial_t h_{\varepsilon,1} &=    2 \varepsilon^{2 -z  -\chi}X (\partial_x h_{\varepsilon,1})(\partial_x h_{\varepsilon,2})
     + \tfrac{1}{2}\varepsilon^{2 -z}T\partial_x^2  h_{\varepsilon,1} +\varepsilon^{{(2\chi -z+1)/2}}\sqrt{T}\xi_1  , \\
	\partial_t h_{\varepsilon,2} &=  \varepsilon^{2 -z  -\chi}Y  (\partial_xh_{\varepsilon,1})^2 + \varepsilon^{2 -z  -\chi}(\partial_x h_{\varepsilon,2})^2 + \tfrac{1}{2} \varepsilon^{2 -z}\partial_x^2 h_{\varepsilon,2}+ \varepsilon^{{(2\chi -z+1)/2}}\xi_2.
    \end{aligned}
\end{equation} 
To have a finite limit, neither zero nor infinity, for $h_{\varepsilon,\alpha}$ requires the sum rule
\begin{equation}
\label{2.2b}
 \chi +z = 2.   
\end{equation}
If the steady state has rapidly decaying correlations, then $\chi = \tfrac{1}{2}$ and hence 
\begin{equation}
\label{2.2c}
 z = \tfrac{3}{2}.   
\end{equation}
The diffusion is order $\varepsilon^{\frac{1}{2}}$ and the noise strength order $\varepsilon^{\frac{1}{4}}$. As a consequence, the mesoscopic scale
is characterized by height\,:\,space\,:\,time having the exponents 
$\tfrac{1}{2}:\tfrac{2}{2}:\tfrac{3}{2}$, which explains the notion of $1:2:3$ scaling.

\textcolor{black}{    For $n=1$ the equations of motion are invariant under Galilei transformations,
which fixes the dynamical exponent $z = \tfrac{3}{2}$ \cite{forster,medina}. However, when tilting Eq. \eqref{1.4}, two sound peaks are generated. Galilean invariance is broken and the sum rule \eqref{2.2b} might no longer be valid, even though indicated by one-loop RG~\cite{1993-ertas-kardar}.   The lack of Galilean invariance was observed already earlier for the conserved version of the one-component KPZ equation. From the  scaling argument as above one would conclude that $\chi +z = 4$ \cite{Krug01041997}.
While this is correct in one-loop approximation,  in two-loop RG one obtains small deviations from this sum rule \cite{JanssenPRL}.}

In this contribution, the dynamical correlator of \eqref{1.4} is numerically simulated for a range of parameters. On the available level of precision we always use $z = \tfrac{3}{2}$. 
\subsection{Stability} 
\label{sec2.2}
 Diffusion and noise is stable but the nonlinearity could generate instabilities. The simplest textbook example reads
 \begin{equation}
 \label{2.3}
\partial_t\phi_1 = \partial_x \phi_2, \quad \partial_t\phi_2 = -\partial_x \phi_1, 
\end{equation}
which amounts to
\begin{equation}
 \label{2.4}
\partial_t^2\phi_1 = -\partial_x^2 \phi_1, \quad \partial_t^2\phi_2 = -\partial_x^2 \phi_2, 
\end{equation}
which is the wave equation with the wrong sign. 

To investigate this issue in our context we start from the general system
\begin{equation}
 \label{2.5}
\partial_t \phi_\alpha + \partial_x J_\alpha(\phi_1,...,\phi_n) = 0,
\end{equation}
which can be written in quasi-linear form as
\begin{equation}
\label{2.6}
\partial_t \phi_\alpha + A_{\alpha\beta}(\vec{\phi}) \partial_x\phi_\beta = 0,
\end{equation}
$\alpha,\beta = 1,\dots,n$. The matrix $A$ is the flux Jacobian, which depends on $\vec{\phi} = (\phi_1,\dots,\phi_n)$.
The system is called hyperbolic, if the eigenvalues of $A$ are real for \textit{all} values of $\vec{\phi}$. 
If the system is not hyperbolic, then there are solutions growing exponentially fast. 
In our case, $n = 2$ and $J_1 = -2X \phi_1\phi_2 $, $J_2 = - Y \phi_1^2 -  \phi_2^2$ resulting in
\begin{equation}
\label{2.7}
A = 
-2\begin{pmatrix}
X\phi_2 & X\phi_1\\
Y\phi_1 & \phi_2
\end{pmatrix}.
\end{equation}
The eigenvalues of $A$ satisfy a quadratic equation with discriminant  
\begin{equation}
\label{2.8}
\Delta = (X-1)^2 \phi_2^2 + 4XY \phi^2_1.
\end{equation}
To have $\Delta \geq 0$ requires 
\begin{equation}
\label{2.9}
XY \geq 0. 
\end{equation}
Therefore  in the quadrants II and IV of the  $X$-$Y$ plane, those defined by $XY<0$, the solutions of \eqref{2.1} exhibit instabilities. Thus only for $XY>0$ we can expect to have a time-stationary measure. As a consequence only quadrants I and III will be considered.

Hyperbolicity is a strong condition, since it has to hold for arbitrary field configurations. There could be interesting phenomena in the unstable regime. In fact, as explained in \cite{weinberger2024}, for polariton-exciton systems $\phi_\alpha$ refers to a phase and hence is bounded by definition. Still the phenomenology in the unstable region is very different from stable KPZ.  


\subsection{Dynamic correlator} 
\label{sec2.2a}
To study scaling functions there is a very wide range of options. In this article we give preference to the time-stationary
two-point correlator. In a simulation one runs the system with periodic boundary conditions until it has reached stationarity. This state is now regarded as random initial conditions and one samples  the dynamic spacetime correlator
\begin{equation}\label{2.10}
S_{\alpha\beta}(x,t) = \langle \phi_\alpha(x,t) \phi_\beta(0,0) \rangle, 
\end{equation}
where $\langle \cdot \rangle$ refers to the average in the time-stationary measure.
The static correlator equals 
\begin{equation}\label{2.11}
C_{\alpha\beta}(x) = \langle \phi_\alpha(x) \phi_\beta(0) \rangle = S_{\alpha\beta}(x,0) 
\end{equation}
and the static susceptibility matrix reads
\begin{equation}\label{2.12}
\mathsfit{C}_{\alpha\beta} = \int_\mathbb{R}dx \langle \phi_\alpha(x) \phi_\beta(0) \rangle. 
\end{equation}
Note that the dynamics is stochastically invariant under the transformation $\phi_1(x,t)$ to $-\phi_1(x,t)$, which yields the simplifications 
\begin{equation}\label{2.13}
S_{12}(x,t) = 0 = S_{21}(x,t), \quad C_{12}(x) = C_{21}(x) = 0,\quad  \mathsfit{C}_{12} = \mathsfit{C}_{21} = 0.
\end{equation}
Therefore we list only the diagonal entries of the matrices $S,C, \mathsfit{C}$.

In the simulations reported in Section 3, we observe good convergence to stationarity and a rapid decay to $0$ of $C_{11}(x)$ and $C_{22}(x)$. The dynamical exponent is well fitted with $z= \tfrac{3}{2}$. To be on the safe side, we assume here a general exponent $z_\alpha >0$. Therefore the scaling hypothesis becomes 
\begin{equation}\label{2.14}
S_{\alpha\alpha}(x,t) \simeq t^{-1/z_\alpha} g_\alpha( t^{-1/z_\alpha}x)
\end{equation}
for sufficiently large $(x,t)$. $g_\alpha$ is the scaling function. In principle $g_\alpha$ depends on $(X,Y,T)$. Hence more explicitly we should write
$g_{\alpha,(X,Y,T)}$.  Now let us consider two parameters $(X,Y,T)$ and $(X',Y',T')$. If $g_{\alpha,(X,Y,T)}$ and $g_{\alpha,(X',Y',T')}$ differ only by a scale factor,
then $(X,Y,T)$ and $(X',Y',T')$ are in the \textit{same} universality class. If this is not the case,  they are in \textit{distinct} classes.  

In general, as discussed in \cite{RDK24}, the correlator $S_{\alpha\beta}$ is a full matrix and so is the respective scaling matrix $g_{\alpha\beta}$.
Two parameter values are in the same universality class, if the corresponding scaling matrices are related by rotation and dilation. 
In our case these matrices are diagonal, which fixes the frame, and thus only dilations have to be considered.
Our investigations are focused to the question of how the three-dimensional $X$-$Y$-$T$ space with $XY >0$, $T>0$ is partitioned into universality classes. 
\subsection{Cole-Hopf transformation}
\label{sec2.3}
As discovered in \cite{1993-ertas-kardar} and more concisely discussed in \cite{FH17}, only upon setting $T=1$, scanning the half-line $\{(1,Y,1),Y>0\}$ allows for a Cole-Hopf transformation and thereby a link to the stochastic heat equation. One starts from the matrix 
\begin{equation}
\label{2.15}
R = 
2\begin{pmatrix}
 \sqrt{Y} & 1\\
 - \sqrt{Y} &1
\end{pmatrix}
\end{equation}
and defines the transformed fields as 
\begin{equation}
\label{2.16}
\tilde{h}  = R h.
\end{equation}
Then the transformed Eq. \eqref{1.4} becomes 
\begin{equation}
\label{2.17}
\partial_t \tilde{h}_ \alpha= \tfrac{1}{2} (\partial_x\tilde{h}_\alpha)^2 +  \tfrac{1}{2} \partial_x^2 \tilde{h}_\alpha+ \tilde{\xi}_\alpha .
\end{equation}  
This looks like magically  the system has been decoupled. But the interaction is hidden in the noise term which is still spacetime Gaussian
white noise but correlated as
\begin{equation}
\label{2.18} 
\langle \tilde{\xi}_\alpha(x,t) \tilde{\xi}_\beta(x',t')\rangle = 4\big((1+Y)\delta_{\alpha\beta} + (1-Y)(1- \delta_{\alpha\beta})\big)\delta(x-x')\delta(t - t'). 
\end{equation} 
For $Y=1$ the system becomes indeed decoupled into two independent KPZ equations, which can be seen also directly from \eqref{1.4}  by a $\pi/4$ rotation. 

The Cole-Hopf transformation is defined by
\begin{equation}
\label{2.19} 
Z_\alpha = \exp(\tilde{h}_\alpha),
\end{equation} 
which yields the two-component stochastic heat equation
\begin{equation}
\label{2.20} 
\partial_t Z_\alpha = \tfrac{1}{2}\partial_x^2 Z_\alpha + \tilde{\xi}_\alpha Z_\alpha.
\end{equation} 
This is the imaginary time Schr\"{o}dinger equation with a random potential. The path integral solution can be written as  a directed polymer moving forward in time and subject to the spacetime random potential $\tilde{\xi}_\alpha(x,t)$.  This potential is correlated,  but the correlations are in force only when the two polymers cross.
Only for $Y=1$ the two components are independent copies. An oversimplified model are two random walks with on-site interaction, which is symmetric under interchange. Then after $t$ steps the number of collisions is of order $\sqrt{t}$. For long times the two random walks become independent. Directed polymers are more complicated objects than random walks, but a corresponding argument still applies. 

To exploit this connection for the dynamic correlator one faces the difficulty that neither the actual nor the transformed time-stationary measure
 is known. A more accessible observable would be to start from flat initial conditions, $h_\alpha(x,0) = 0$. Then the transformed initial conditions are also flat and one arrives at two point-to-line directed polymers. Still the crossings occur  
only for a negligible fraction of the total time $t$. Therefore the height fluctuations of $\tilde{h}_\alpha$ should converge to two independent GOE Tracy-Widom distributions, denoted by here $\vartheta_\alpha$. Inverting $R$, up to constant multiplicative  factors, the height fluctuations of $h_1$ are given by the sum  
$(\vartheta_1 + \vartheta_2)/\sqrt{2}$ and of  $h_2$ by the difference $(\vartheta_1 - \vartheta_2)/\sqrt{2}$. But these are exactly the height fluctuations at $Y=1$. Thus the Cole-Hopf line $\{(1,Y,1),Y>0\}$  is a single universality class with fixed point $(1,1,1)$.
This conclusion is further confirmed 
by \cite{weinberger2024}, where direct numerical simulations  are reported at $(1,2,1)$ for flat initial conditions and independent GOE statistics is observed. In addition, in Section \ref{sec3} we report on numerical simulations of the dynamic correlator at $(0.4,1,1)$ and $(1,1.2,1)$.
\subsection{Cyclicity and fixed points}
\label{sec2.5}
For a short moment we return to the general case \eqref{1.3}, setting $n=2$. As noted in \cite{2014-spohn} and further elaborated by \cite{FH17,FH15,H23} for a particular parameter choice one can still compute the time-stationary measure.  
The linear part has a Gaussian stationary measure with covariance $C_{\alpha\beta}(x) = \delta(x)\mathsfit{C}_{\alpha\beta}$,
where $\mathsfit{C}$
 is the solution of
\begin{equation}
\label{2.25a}
\tfrac{1}{2}\big(D\mathsfit{C} + \mathsfit{C}D^\mathrm{T}\big) = BB^\mathrm{T}
\end{equation}
with $(\cdot)^\mathrm{T}$ denoting the transpose of a matrix.
We define
\begin{equation}
\label{2.25b}
 \hat{G}^\alpha = (\mathsfit{C}^{-1} )_{\alpha\beta} G^\beta.   
\end{equation}
Then $G^\alpha$ are called \textit{cyclic} with respect to $\mathsfit{C}$, if 
they are of the form 
\begin{equation}
    \label{2.25c} 
    \hat{G}^1 = 
    \begin{pmatrix}
        a & b\\
        b & c\\
    \end{pmatrix}, \qquad 
    \hat{G}^2 = 
    \begin{pmatrix}
        b & c\\
        c & d\\
    \end{pmatrix}
\end{equation}
with arbitrary coefficients $a,b,c,d$. If $G^\alpha$ are cyclic, the full nonlinear equation \eqref{1.3} has the same invariant measure as the linear part of the equation.

In our context the specific choice of matrices is
\begin{equation}
    \label{2.25d} 
    G^1 = 
    \begin{pmatrix}
        0 & X\\
        X & 0\\
    \end{pmatrix}, \quad 
    G^2 = 
    \begin{pmatrix}
        Y & 0\\
        0 & 1\\
    \end{pmatrix},\quad
    \mathsfit{C} = 
    \begin{pmatrix}
        c_1 & 0\\
        0 & c_2\\
    \end{pmatrix}.
\end{equation}
Then the  matrices $G^\alpha$ are cyclic with respect to $\mathsfit{C}$, if and only if
\begin{equation}
\label{2.25e}
c_1Y = c_2X.
\end{equation}
\textcolor{black}{On a formal level, this property can be easily demonstrated.
The white noise has the density
\begin{equation}
\label{2.25ff}
\exp\big[-\tfrac{1}{2}\int dx\big(c_1^{-1}\phi_1^2(x) + c_2^{-1}\phi_2^2(x)\big)\big].
\end{equation}
By construction, this Gaussian measure is stationary under the linear part of the coupled Burgers equations. To ensure stationarity under the full dynamics only the nonlinear part has to checked, which leads to the condition
\begin{equation}
\label{2.25g}
\frac{d}{dt}\int dx\tfrac{1}{2}\big(c_1^{-1}\phi_1^2 + c_2^{-1}\phi_2^2\big)
= \int dx\big(c_1^{-1}2X\phi_1\partial_x(\phi_1\phi_2) + c_2^{-1}\phi_2\partial_x(Y \phi_1^2 + \phi_2^2\big) = 0.
\end{equation}
Denoting the left hand term by $I$, by partial integration one obtains
\begin{equation}
\label{2.25h}
I = \int dx\big(2(-c_1^{-1}X + c_2^{-1}Y)\phi_1\phi_2\partial_x\phi_1 + c_2^{-1}\tfrac{1}{3}\partial_x\phi_2^3\big).
\end{equation}
Clearly $I=0$, if and only if Eq. \eqref{2.25e} holds.}
\begin{figure}[!t]
\begin{center}
    \begin{subfigure}{0.45\linewidth}
	\includegraphics[width=\linewidth]{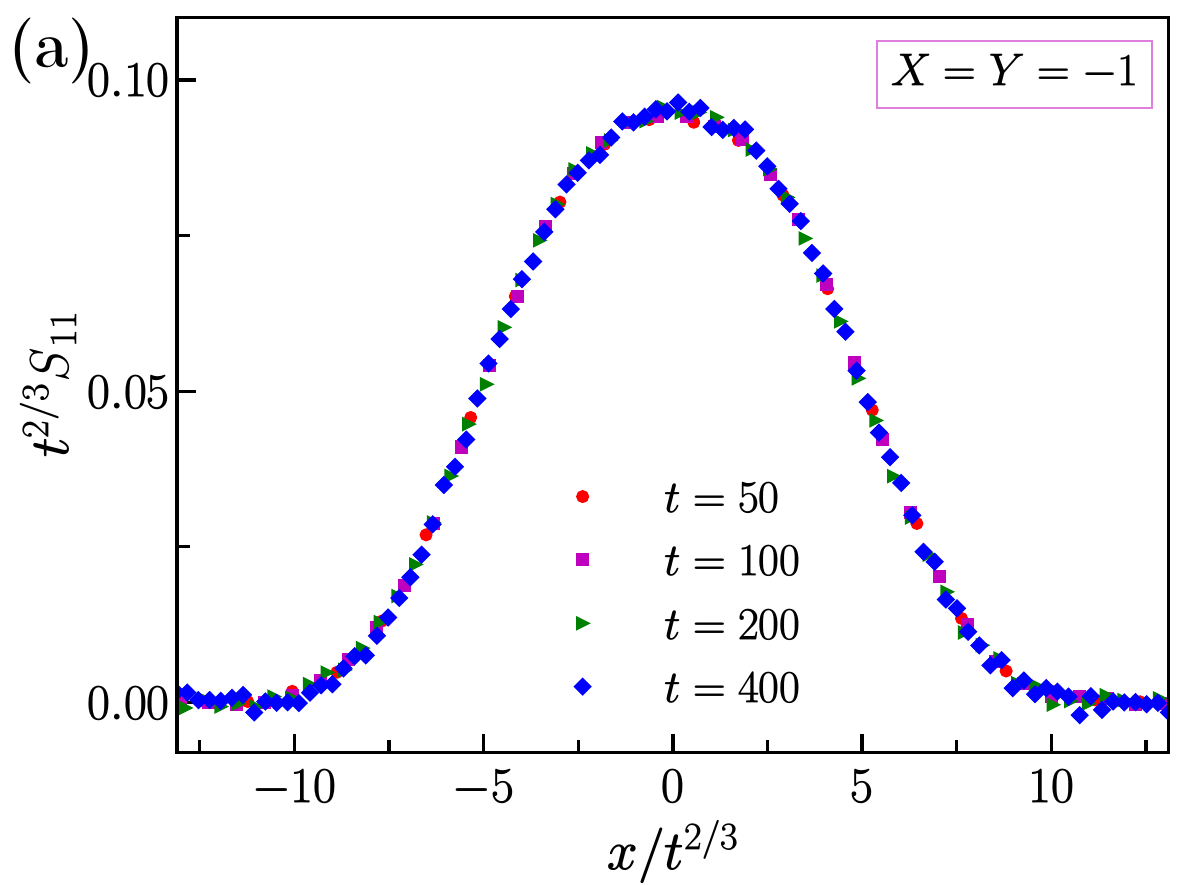}
    \end{subfigure}%
    \begin{subfigure}{0.45\linewidth}
	\includegraphics[width=\linewidth]{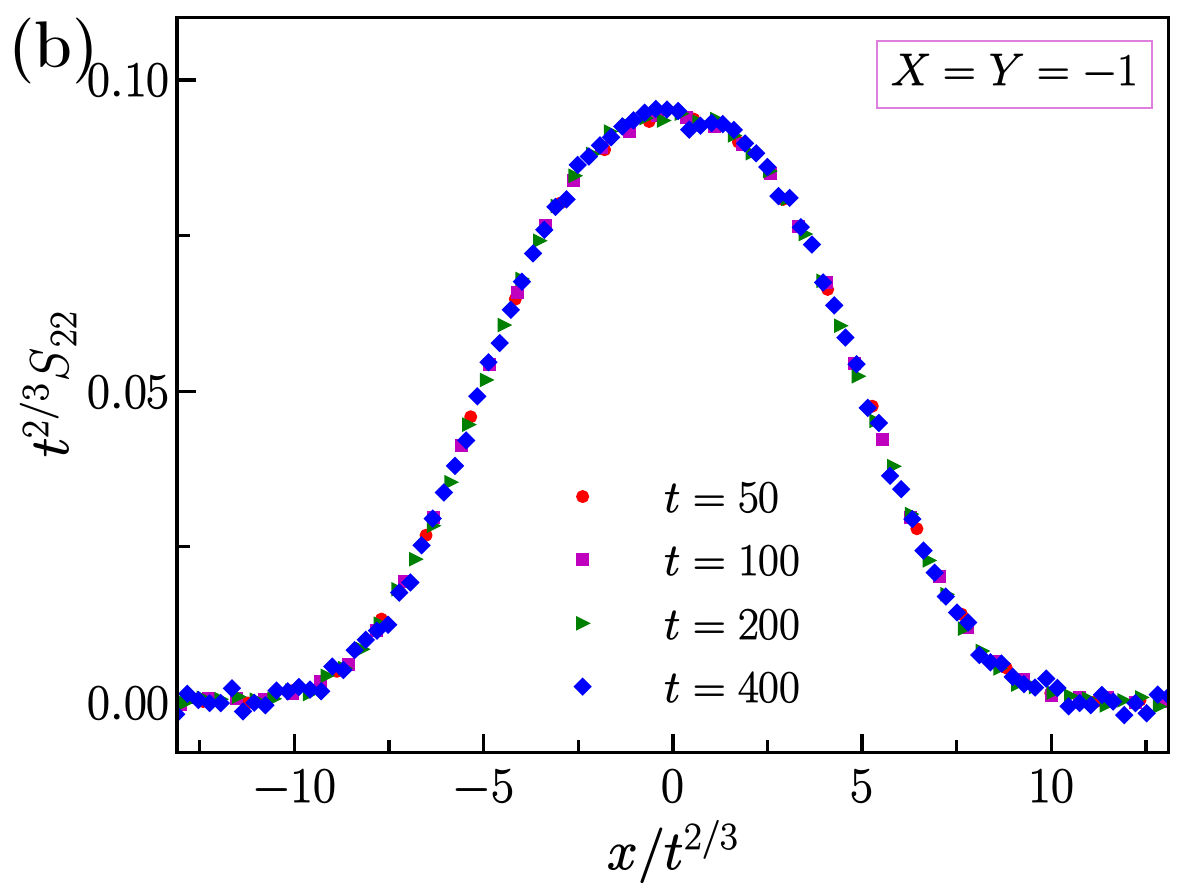}
    \end{subfigure}
    \begin{subfigure}{0.45\linewidth}
	\includegraphics[width=\linewidth]{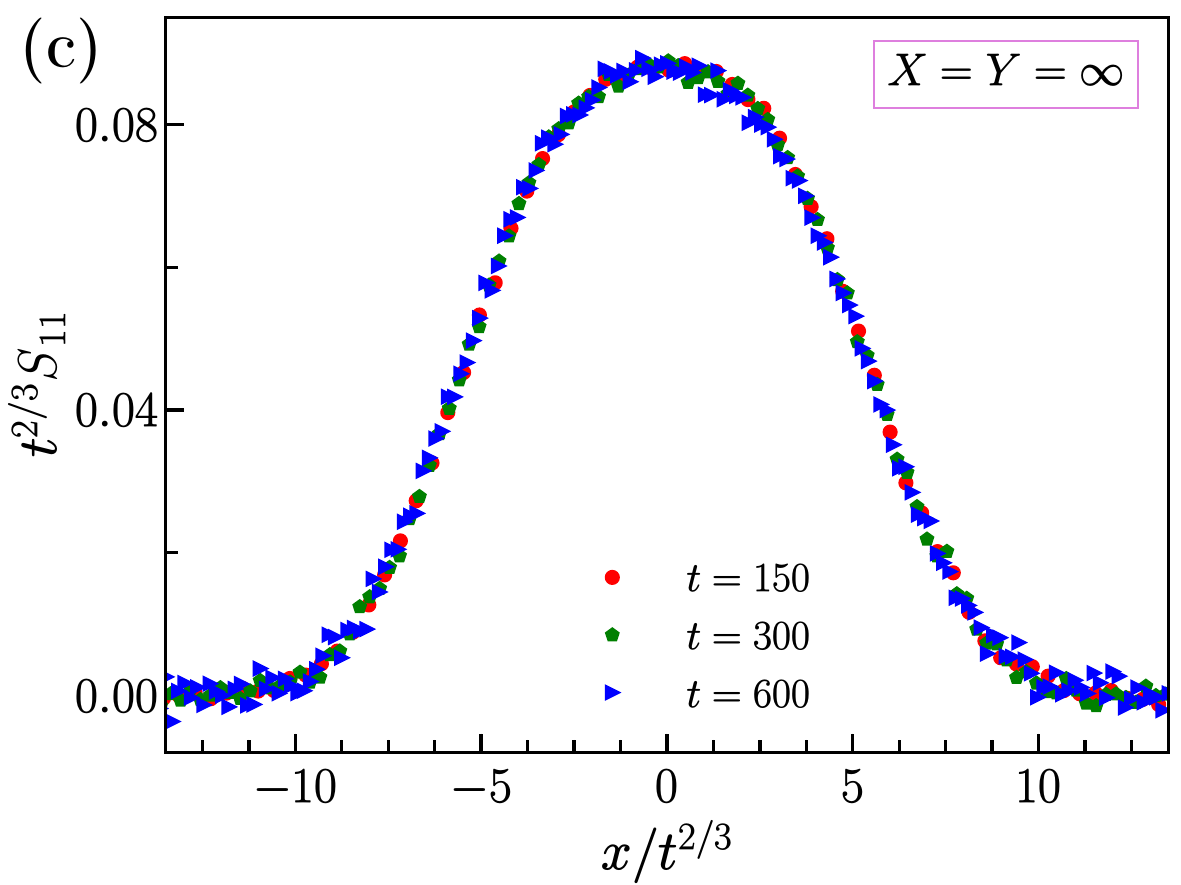}
    \end{subfigure}%
    \begin{subfigure}{0.45\linewidth}
	\includegraphics[width=\linewidth]{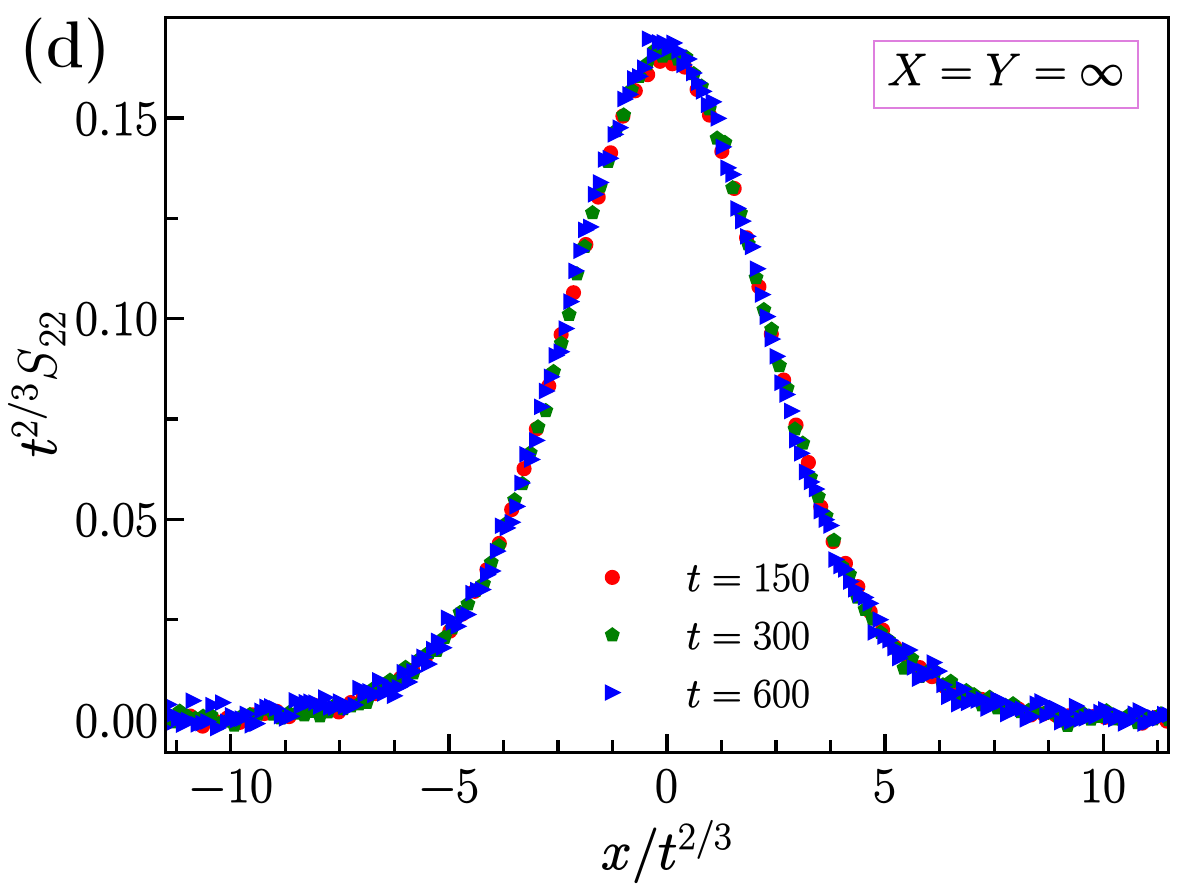}
    \end{subfigure}
    \caption{Spacetime correlations for $X=Y=-1$ and $X=Y\to\infty$, setting $T=1$. For $(-1,-1,1)$, plotted are the rescaled spacetime correlations (a) $S_{11}$ and (b) $S_{22}$ at different times $t=50, 100, 200, 400$. For $(\infty,\infty,1)$ plotted are rescaled spacetime correlations (a) $S_{11}$ and (b) $S_{22}$ at different times $t=150, 300, 600$. The simulation is based on Eq. \eqref{3.1}, see below. We set $\mathsfit{b}=3$ and average over $10^5$ independent realizations. The observed dynamic scaling exponent is $\approx 0.67$.}
    \label{fig:a}
\end{center}
\end{figure}

Imposing cyclicity and normalizing to $\mathsfit{C}=1$,
the coupled Burgers equations read
\begin{equation}\label{2.25f}
    \begin{aligned}
	\partial_t \phi_1 &= \partial_x \big(  2 X  \phi_1\phi_2 + \tfrac{1}{2}T \partial_x \phi_1+ \sqrt{T}\xi_1 \big), \\
	\partial_t \phi_2 &= \partial_x \big( X  \phi_1^2 + \phi_2^2 +  \tfrac{1}{2}\partial_x \phi_2  +\xi_2 \big).
    \end{aligned}
\end{equation}  
The scaling functions  depend on $(X,X,T)$. Their decomposition into universality classes has still to be figured out.
In \cite{RDK24} we tried to clarify this issue by direct numerical simulations in case $T=1$. 
For the ease of the reader, in Fig.~\ref{fig:a} we display the scaled dynamic correlator for the parameters $(-1,-1,1)$  and also for $(\infty,\infty,1)$, which is the limit $X \to \infty$ under the constraint $Y=X$, see Section \ref{sec2.8}.  In Fig.~\ref{fig:a} we have plotted  the scaling functions  $g_{1,(X,X,1)}$ and
$g_{2,(X,X,1)}$, which are referred to as peak 1 and 2. At $X=1$ one observes that both peaks are equal to the exact KPZ scaling function. As $X$ is increased, peak 1  becomes fatter and  peak 2 taller under constraint of having area $1$ under the peak. The most pronounced  anisotropy occurs at $X=\infty$, at which the ratio (peak top 2/peak top 1) $ \simeq 2$. Near the top, peak 1 is considerably flatter  than the KPZ peak. For $0 <X< 1$ the order of peak heights is reversed. For $X< 0$, we have only a single data point at $X=-1$, at which the peaks have equal height. Qualitatively the    
order of peaks is reflected at the $Y$ axis, but there is no strict reflection symmetry.

Variations in  $X$ result in moderate changes of the scaling functions. Further support comes from the cyclic parameter points
$(1,1,1)$ in Figure \ref{fig:cb1} and $(2,2,1)$ in Figure \ref{fig:cb2}. Taking all available evidence into account, we conclude that $g_{1,(X,X,1)}$ and $g_{2,(X,X,1)}$ are non-trivially modified when scanning $X$. Each one of them thus constitutes a distinct universality class. 

For this study, in addition we investigate the dependence on $T$ and simulated the parameter points $(2,2,0.5)$ and $(2,2,2)$. As discussed in Section 3, the value of $T$  results only in non-universal coefficients. Hence we 
expect that a universality class consists precisely of a half-plane parallel to the $Y$-$T$-plane passing  through the point $(X,X,1)$.
\subsection{The full phase diagram}
\label{sec2.6}

 The task is to study the coupled noncyclic Burgers equation \eqref{2.1}, 
 \begin{equation}\label{2.1a}
    \begin{aligned}
	\partial_t \phi_1 &= \partial_x \big(  2 X  \phi_1 \phi_2 + \tfrac{1}{2} T\partial_x \phi_1+ \sqrt{T}\xi_1 \big), \\
	\partial_t \phi_2 &= \partial_x \big(Y  \phi_1^2 + \phi_2^2 +  \tfrac{1}{2}\partial_x \phi_2  + \xi_2 \big).
    \end{aligned}
\end{equation}
The steady state is no longer $\delta$-correlated and 
\begin{equation}\label{2.1e}
\langle \phi_\alpha(x)\phi_\alpha(x')\rangle = C_{\alpha\alpha}(x-x').
\end{equation}
As our standing assumption, confirmed by simulations, $C_{\alpha\alpha}(x)$ decays rapidly and has a finite correlation length, denoted by $\eta$. In particular, the susceptibility 
\begin{equation}\label{2.1b}
\int_\mathbb{R}dx \,C_{\alpha\alpha}(x) = c_\alpha >0.
\end{equation}
As a precursor of the mesoscopic scale, the field $\phi_\alpha$ is coarse-grained over a scale much larger than $\eta$. Denoting the coarse-grained field by $\tilde{\phi}_\alpha$, this results in  
\begin{equation}\label{2.1d}
\langle \tilde{\phi}_\alpha(x)\tilde{\phi}_\alpha(x')\rangle = c_\alpha\delta(x-x').
\end{equation}

As a next step we want to write down an effective coupled KPZ equation from which the universality classes can be deduced. Universal scaling properties are captured by the $1:2:3$ scaling, see Section \ref{sec2.1}. Thereby the solution $\phi_\alpha(x,t)$ consists of two widely separated parts, a microscopic and mesoscopic part: the latter is a smooth envelope function of order $1$, still random, which is perturbed by in essence white noise of strength $\varepsilon^\frac{1}{4}$. With this reasoning, the operations of coarse-graining and forming products should commute, 
\begin{equation}\label{2.1h}
(\phi_\alpha \phi_\beta)\,\tilde{} =  \tilde{\phi}_\alpha \tilde{\phi}_\beta, 
\end{equation}
up to small errors.

For a patch on the microscopic scale, the nonlinearites can be regarded as being constant and only diffusion plus noise has to  be handled.
To account for a finite correlation length we construct a phenomenological Gaussian model, for which purpose it is convenient to use Fourier space.
Then the diffusion constant, $D_\alpha(k)$, is taken as $k$-dependent, while the noise strength is a constant $\sigma_\alpha$. In Fourier space the Langevin equation is then given by
\begin{equation}
\label{2.26a}
\partial_t \hat{\phi}_\alpha(k,t) =   -\tfrac{1}{2}k^2D_\alpha(k)\sigma_\alpha^2\hat{\phi}_\alpha(k,t) + \mathrm{i}k \sigma_\alpha\hat{\xi}_\alpha(k,t), 
\end{equation}
where $\sigma_1 = \sqrt{T}$ and $\sigma_2 = 1$ in the case under study.
The Fourier transform of the static correlator is $\hat{C}_{\alpha\alpha}(k) = 1/D_\alpha(k)$. In real space, $C_{\alpha\alpha}(x)$ decays rapidly and hence $D(k)$ diverges at large $k$ thereby suppressing the corresponding Fourier modes. The susceptibility is $c_\alpha = \hat{C}_{\alpha\alpha}(0) = 1/D_\alpha(0)$. Coarse-graining over scales much larger than $\eta$ amounts to discarding the large $k$ behavior. Switching back to physical space and denoting the coarse-grained fields by $\tilde{\phi}_\alpha$ yields the effective Langevin equation
\begin{equation}
\label{2.26b}
\partial_t \tilde{\phi}_\alpha(x,t) =   \tfrac{1}{2}c_\alpha^{-1}\sigma_\alpha^2\partial_x^2\tilde{\phi}_\alpha(x,t) + \sigma_\alpha\partial_x\xi_\alpha(x,t). 
\end{equation}
In our approximation spatial coarse-graining merely modifies the diffusion constant.

Adding, after coarse graining, the nonlinear terms present in \eqref{2.1a}  back into \eqref{2.26b} leads to
\begin{equation}\label{2.1g}
    \begin{aligned}
	\partial_t \tilde{\phi}_1 &= \partial_x \big(  2 X  \tilde{\phi}_1 \tilde{\phi}_2 + \tfrac{1}{2} c_1^{-1}T\partial_x \tilde{\phi}_1+ \sqrt{T}\xi_1 \big), \\
	\partial_t \tilde{\phi}_2 &= \partial_x \big(Y  \tilde{\phi}_1^2 + \tilde{\phi}_2^2 +  \tfrac{1}{2}c_2^{-1}\partial_x \tilde{\phi}_2  + \xi_2 \big).
    \end{aligned}
\end{equation}

We now return to the true steady state obtained from \eqref{2.1a} and recall that under coarse-graining  the correlator becomes $c_\alpha \delta(x-x')$, see \eqref{2.1e} and \eqref{2.1d}. However  a much stronger property holds generically for systems away from criticality. Under coarse-graining, the steady state field $\tilde{\phi}_\alpha(x)$ is actually Gaussian white noise of strength $c_\alpha$. The two components are independent. We have obtained two properties based on disjoint arguments, the effective evolution equation \eqref{2.1g} and the white noise statistics of the steady state. According to the discussion in Section \ref{sec2.5}, Gaussian white noise as stationary measure necessarily implies that the model is cyclic relative to $\mathsfit{C}_{\alpha\beta} = c_\alpha\delta_{\alpha\beta} $ and hence
\begin{equation}
\label{2.28a}
c_2X = c_1 Y.
\end{equation}

Our result becomes more transparent through carrying out standard rescalings. First   the fields are normalized as
\begin{equation}
 \label{2.26e}
 \check{\phi}_\alpha= \frac{1}{\sqrt{c_\alpha}} \tilde{\phi}_\alpha
\end{equation}
and the  relation \eqref{2.28a} is implemented. Then the dynamics of the new fields is governed by
\begin{equation}
 \label{2.26g}
    \begin{aligned}
        \partial_t \check{\phi}_1 &= \sqrt{c_2}\partial_x \big(  2 X\check{\phi}_1 \check{\phi}_2 \big) +\partial_x\big(\tfrac{1}{2}c_1^{-1}T\partial_x \check{\phi}_1 + c_1^{-1/2}\sqrt{T}\xi_1\big) , \\
        \partial_t \check{\phi}_2 &= \sqrt{c_2}\partial_x \big(X\check{\phi}_1^ 2 +  \check{\phi}_2^2\big) + \partial_x\big(\tfrac{1}{2}c_2^{-1}\partial_x  \check{\phi}_2 + c_2^{-1/2}\xi_2\big). \end{aligned}
\end{equation}
A further rescaling, as in Appendix A, with $a_1=a_2=c_2^{-3/2}, \ell=c_2^{-3},~\tau=c_2^{-5}$ leads to the equivalent equation
\begin{equation}
 \label{2.26g:sc}
    \begin{aligned}
        \partial_t \check{\phi}_1 &= \partial_x \big(  2 X\check{\phi}_1 \check{\phi}_2 \big) +\partial_x\big(\tfrac{1}{2}T_{\rm eff}\partial_x \check{\phi}_1 + \sqrt{T_{\rm eff}}\xi_1\big) , \\
        \partial_t \check{\phi}_2 &= \partial_x \big(X\check{\phi}_1^ 2 +  \check{\phi}_2^2\big) + \partial_x\big(\tfrac{1}{2}\partial_x  \check{\phi}_2 + \xi_2\big), \end{aligned}
\end{equation}
where $T_{\rm eff}=(c_2/c_1)T$. 
It then follows that the bare parameters $(X,Y,T)$ are in the same universality class 
as the  point $(X,X,T_{\rm eff})$. 

Physically one would expect that the large scale behavior depends only on the susceptibility, but not separately on diffusion and noise strength. 
Thus, the points  $(X,X,T)$ with arbitrary $T$ should be in same universality class, which is confirmed by the simulations to be discussed in Section \ref{sec3.3}. By convention, we can regard $(X,X,1)$ as the fixed point defining the class.
Then our final conclusion is that $(X,Y,T)$ is in the universality class characterized by the fixed point $(X,X,1)$.
This feature is partially visualized in Figure \ref{fig1} by arrows pointing from the bare $(X,Y)$ towards the fixed point
$(X,X)$.

Somewhat unexpectedly, our argument yields the novel relation \eqref{2.28a} which involves only 
properties of the steady state. Thereby we have acquired  a simple test whether our reasoning is valid. Significant deviations would mean that our arguments require further improvements.

\subsection{Limiting parameters}
\label{sec2.8}
Throughout this sub-section we set $T= 1$ and discuss the limiting parameters $(0,0,1)$, $(0,Y,1)$, and $(X,0,1)$. The fourth case is the limit $X\to \infty$ with $Y = \kappa X$, $\kappa > 0$.\\\\
\textit{(0,0)}: The two components are uncoupled and satisfy a linear Langevin equation, which is referred to as Gaussian universality. \\\\
 \textit{(0,Y)}: The equations of motion are
 \begin{equation}\label{2.35}
    \begin{aligned}
	\partial_t \phi_1 &= \partial_x \big(  \tfrac{1}{2} \partial_x \phi_1+ \xi_1 \big), \\
	\partial_t \phi_2 &= \partial_x \big( Y  \phi_1^2 + \phi_2^2 +  \tfrac{1}{2}\partial_x \phi_2  +\xi_2 \big).
    \end{aligned}
\end{equation} 
Clearly component 1 is Gaussian. For component 2 the noise  $\xi_2$ is modified to $Y\phi_1^2  + \xi_2$, where 
the two terms are independent.
The truncated spacetime covariance of $\phi_1^2$ reads
\begin{equation}
\label{2.36}
\langle \phi_1(x,t)^2 \phi_1(0,0)^2 \rangle^\mathrm{c} = 2\langle \phi_1(x,t) \phi_1(0,0) \rangle^2 = \frac{1}{\pi t}\exp(- x^2/t).
\end{equation}  
This term is decaying to zero for long times. Thus the asymptotics of the $\phi_2$ field is expected to satisfy KPZ scaling.\\\\
\textit{(X,0)}: The equations of motion are
\begin{equation}\label{2.37}
    \begin{aligned}
	\partial_t \phi_1 &= \partial_x \big(  2 X  \phi_1 \phi_2 + \tfrac{1}{2} \partial_x \phi_1+ \xi_1 \big), \\
	\partial_t \phi_2 &= \partial_x \big( \phi_2^2 +  \tfrac{1}{2}\partial_x \phi_2  +\xi_2 \big),
    \end{aligned}
\end{equation}
The $\phi_2$ field  is decoupled and satisfies KPZ scaling. The feedback to the field $\phi_1$ is more tricky.
It is still a linear Langevin equation, but there is a linear drift
with an independent spacetime stationary random strength.  The average $\langle \phi_1(x,t) \rangle =0$.
This indicates that the $\phi_1$ field converges to a Gaussian, as in case $(0,0)$.\\\\
$(X,\kappa X), \,\,large\,\, \kappa$: Setting $Y = \kappa X$, the common factor $X$ of the nonlinearity can be absorbed by rescaling time. The coefficient in front of $\phi_2^2$ equals $1/X$.
Hence in the limit $X \to \infty$, the equations read
\begin{equation}\label{2.38}
    \begin{aligned}
	\partial_t \phi_1 &= \partial_x \big(  2 \phi_1 \phi_2 + \tfrac{1}{2} \partial_x \phi_1+ \xi_1 \big), \\
	\partial_t \phi_2 &= \partial_x \big( \kappa  \phi_1^2 +  \tfrac{1}{2}\partial_x \phi_2  +\xi_2 \big),
    \end{aligned}
\end{equation} 
which should be viewed as a coupled stochastic Burgers equation in its own right. For $\kappa = 1$, the model is cyclic and hence the time-stationary measure consists of two independent 
standard spatial white noises. In fact, $\kappa =1$ has been studied already in 
\cite{RDK24} corresponding to the case with parameter $\lambda = 0$.  It is for this choice of parameters that one observes the yet strongest deviations from the 
KPZ scaling function, compare with  Fig. \ref{fig:a}.  Away from this fixed point the time-stationary measure is not known. However \eqref{2.28a} turns 
into the prediction
\begin{equation}
\label{2.39}
 \frac{c_2}{c_1} = \kappa.
\end{equation} 
For all values of $\kappa$  one expects the scaling functions to converge asymptotically to  the ones of the fixed point $\kappa = 1$. From the viewpoint of our theory, Eq. \eqref{2.38} has a minimal number of parameters and thus would be an interesting testing ground.
\section{Direct numerical simulations}
\label{sec3}
\setcounter{equation}{0}
\subsection{Static and dynamic correlators}
\label{sec3.1}
We simulated the coupled Burgers equations 
\begin{equation}
\label{3.1}
    \begin{aligned}
	\partial_t \phi_1 &= \partial_x \big(2 \mathsfit{b} X  \phi_1 \phi_2 + \tfrac{1}{2}\mathsfit{d}T\partial_x \phi_1 +\sqrt{\mathsfit{d}T}\xi_1 \big), \\
	\partial_t \phi_2 &= \partial_x \big(\mathsfit{b}Y  \phi_1^2 + \mathsfit{b} \, \phi_2^2 +\tfrac{1}{2}\mathsfit{d}\partial_x \phi_2 + \sqrt{\mathsfit{d}}\xi_2 \big).
    \end{aligned}
\end{equation}  
We introduced a parameter $\mathsfit{b}$ which controls the strength of the nonlinear term and likewise $\mathsfit{d}$ for the linear part. Note that $\mathsfit{b} =1$ and 
$\mathsfit{d} =2$ reduces to \eqref{2.1}.
Varying either $\mathsfit{b}$ or $\mathsfit{d}$, one stays in the same universality class, as can be deduced from the identities 
\eqref{A.6} and \eqref{A.7}. In our simulations we always set 
$\mathsfit{d} = 2$. While the parameter $\mathsfit{b}$ looks arbitrary, for numerical simulations on a finite grid, there is an optimal window for its choice. 
 We first consider $T=1$. The values of $\mathsfit{b}$
are listed in Table 1 below. Secondly we vary $T$ keeping $X=2$ fixed. In this case the good choice is $\mathsfit{b} = 2$.

The technical details of the simulation are provided in \cite{RDK24}. The noncyclic case requires however additional considerations.

The finite grid consists of  the integers $ 1 \leq j \leq L $  with periodic boundary conditions, where $L = 2048$
for our simulations. To distinguish from the continuum, 
the fields are denoted by $\phi_{\alpha,j} (t)$. In the initial measure the fields $\phi_{\alpha,j} (0)$ are i.i.d. Gaussians with mean $0$ and variance $1$. When $X=Y$, this is already the steady state and one samples the dynamic correlator as 
\begin{equation}
\label{3.2}
S_{\alpha\alpha}(j-i,t) = \langle \phi_{\alpha,j}(t)\phi_{\alpha,i} (0)\rangle.
\end{equation} 
In our simulations the number of independent samples is of the order $10^4$. 

However when $X\neq Y$, the situation is slightly more involved. First one has to equilibrate the system,  through running the dynamics up to some sufficiently long time, here denoted by $t_\mathrm{eq}$. We tested equilibration times from $t_\mathrm{eq}=500$ to $t_\mathrm{eq} = 5000$. As a check of stationarity, considered is
the equal time correlator
\begin{equation}
\label{3.3}
C_{\alpha\alpha}(j-i,t) = \langle \phi_{\alpha,j}(t)\phi_{\alpha,i} (t)\rangle.
\end{equation} 
For $t >t_\mathrm{eq}$ this average should remain unchanged. Applying  standard rules blindly, the static susceptibility is the
defined by 
\begin{equation}
\label{3.4}
\sum_{j=1}^{L} C_{\alpha\alpha}(j,t) = \chi_{\alpha}(t).
\end{equation}
Since Eq. \eqref{3.1} is of conservation type, $\chi_{\alpha}(t)$ does not depend on time and hence  $\chi_{\alpha}(t) =1$. In actual fact, $C_{\alpha\alpha}(j,t)$ has a peak centered at the origin 
with width of order $1$ and
a constant background
of amplitude $1/L$. This can be seen more clearly by considering the Fourier transform 
\begin{equation}
\label{3.5}
\hat{C}_{\alpha\alpha}(k,t_\mathrm{eq}) = \sum_{j=1}^L\mathrm{e}^{\mathrm{i}kj}C_{\alpha\alpha}(j,t_\mathrm{eq}),
\end{equation}
where $k = (2\pi/L)m$, $m= 0,...,L-1$.
Then $\hat{C}_{\alpha\alpha}(0,t_\mathrm{eq}) = 1$. But the physical susceptibility 
equals
 \begin{equation}
\label{3.6}
c_\alpha =
\lim_{k\to 0}\hat{C}_{\alpha\alpha}(k,t_\mathrm{eq}) . 
\end{equation}
Of course, $k$ runs only over a grid with spacing $2\pi/L$. So the limit \eqref{3.6} is understood in sense of a quadratic fit close to $k=0$. More precisely, for $k \neq 0$, but close to $0$, one uses the fit function
$f_\mathrm{fit}(k) = c_\alpha - ak^2$, $a >0$.

The proper definition of the dynamic correlator is then
\begin{equation}
\label{3.7}
S_{\alpha\alpha}(j-i,t) = \langle \phi_{\alpha,j}(t_\mathrm{eq} +t)\phi_{\alpha,i} (t_\mathrm{eq})\rangle +\tfrac{1}{L}(-1 +c_\alpha).
\end{equation}   
Summing $S_{\alpha\alpha}(j-i,t)$ over $j$, the first summand results in $1$ and both summands add up to $c_\alpha$, as it should be. Finally, one notes that upon setting
\begin{equation}
\label{3.8}
\hat{\phi}_{\alpha}(k,t) = \frac{1}{\sqrt{L}}\sum_{j=1}^L\mathrm{e}^{\mathrm{i}kj}\phi_{\alpha,j}(t),
\end{equation}
the Fourier transform of the equal time correlator reads
\begin{equation}
\label{3.9}
\langle|\hat{\phi}_{\alpha}(k,t)|^2\rangle  = \hat{C}_{\alpha\alpha}(k,t).
\end{equation}

As a routine control we confirm that the off-diagonal matrix elements of $C_{\alpha\beta}(j,t)$ and $S_{\alpha\beta}(j,t)$ indeed vanish within statistical error bars.

In the introduction, see Figure \ref{fig1}, we displayed the parameter points for which simulations have been carried out. Out of those, the Table below lists two blocks, $X=1$ and $X=2$, each consisting of three parameter points. One is the fixed point, the two others are above and below the fixed point, all at the same value of $X$. 
The parameter $\mathsfit{b}$ refers to \eqref{3.1} and $p_\alpha$ is the maximum height of peak $\alpha$. According to theory, the ratios $c_2p_1 / c_1 p_2$ and $c_2 X/c_1 Y$
should be independent of $Y$ for given X.

\begin{center}
\begin{figure}[ht]
            \begin{subfigure}{0.33\linewidth}
			\includegraphics[width=\linewidth]{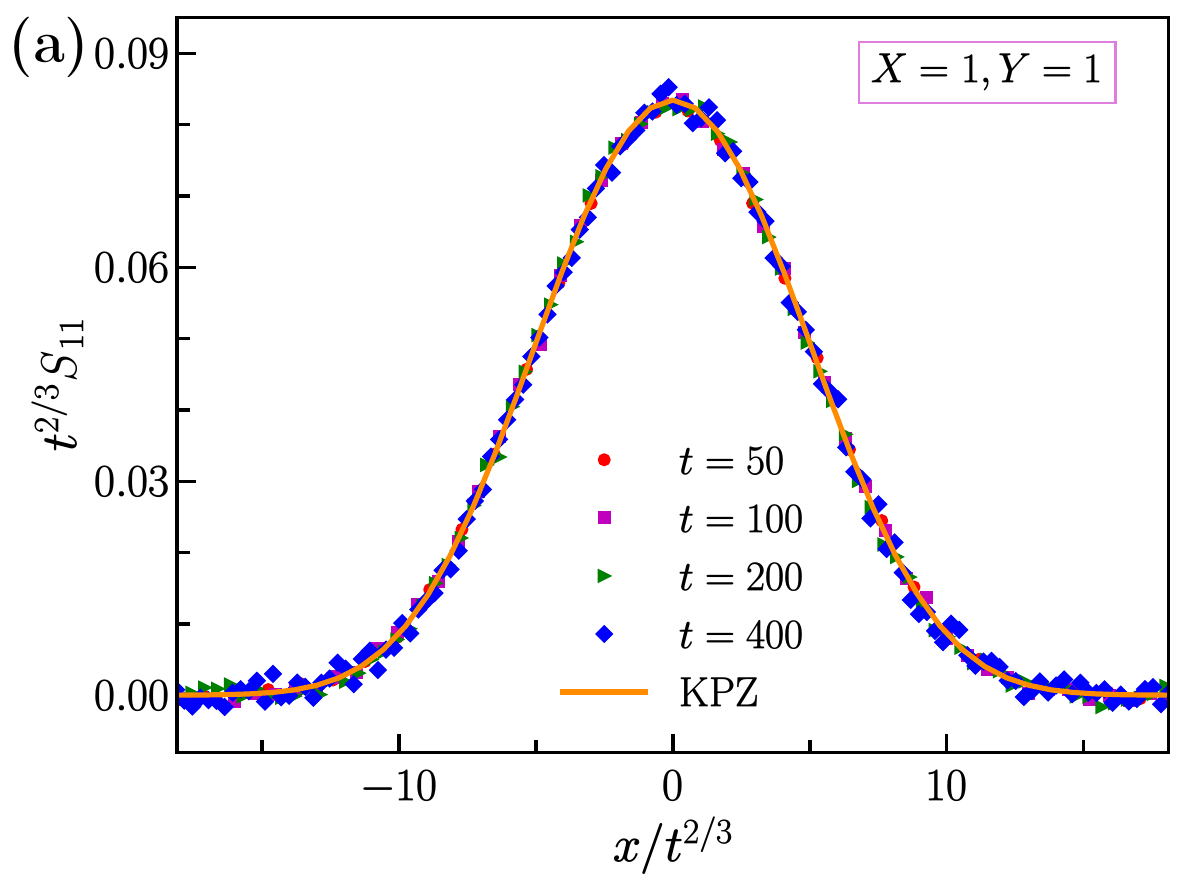}
		\end{subfigure}%
		\begin{subfigure}{0.33\linewidth}
			\includegraphics[width=\linewidth]{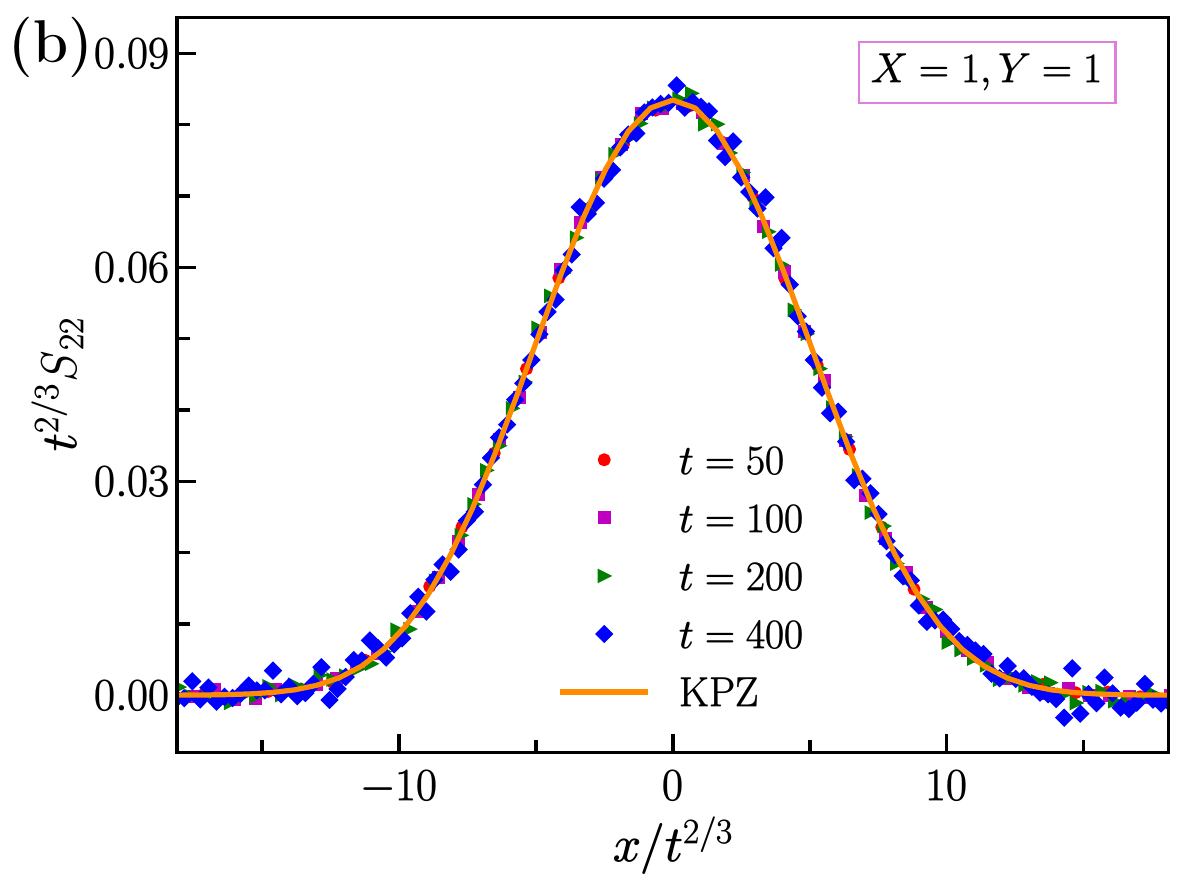}
		\end{subfigure}%
		\begin{subfigure}{0.33\linewidth}
			\includegraphics[width=\linewidth]{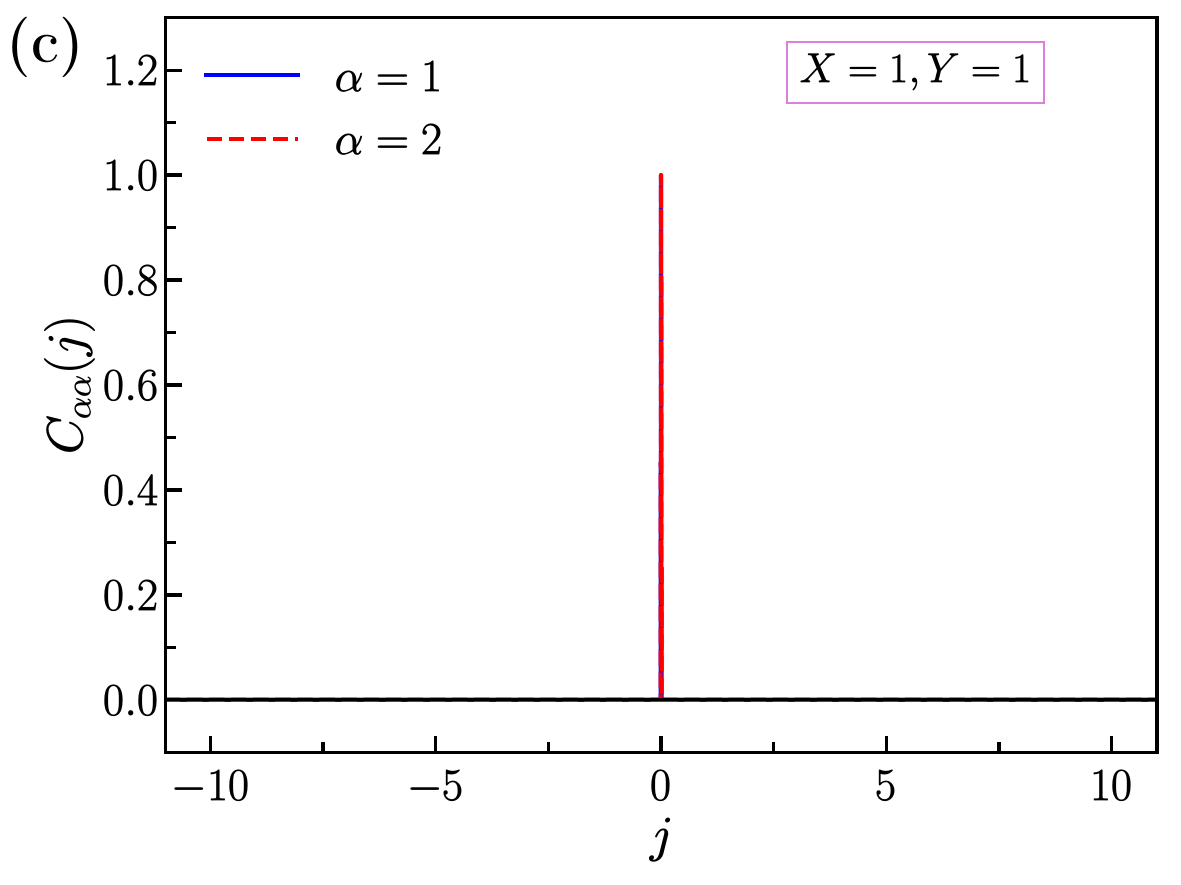}
		\end{subfigure}
            \begin{subfigure}{0.33\linewidth}
			\includegraphics[width=\linewidth]{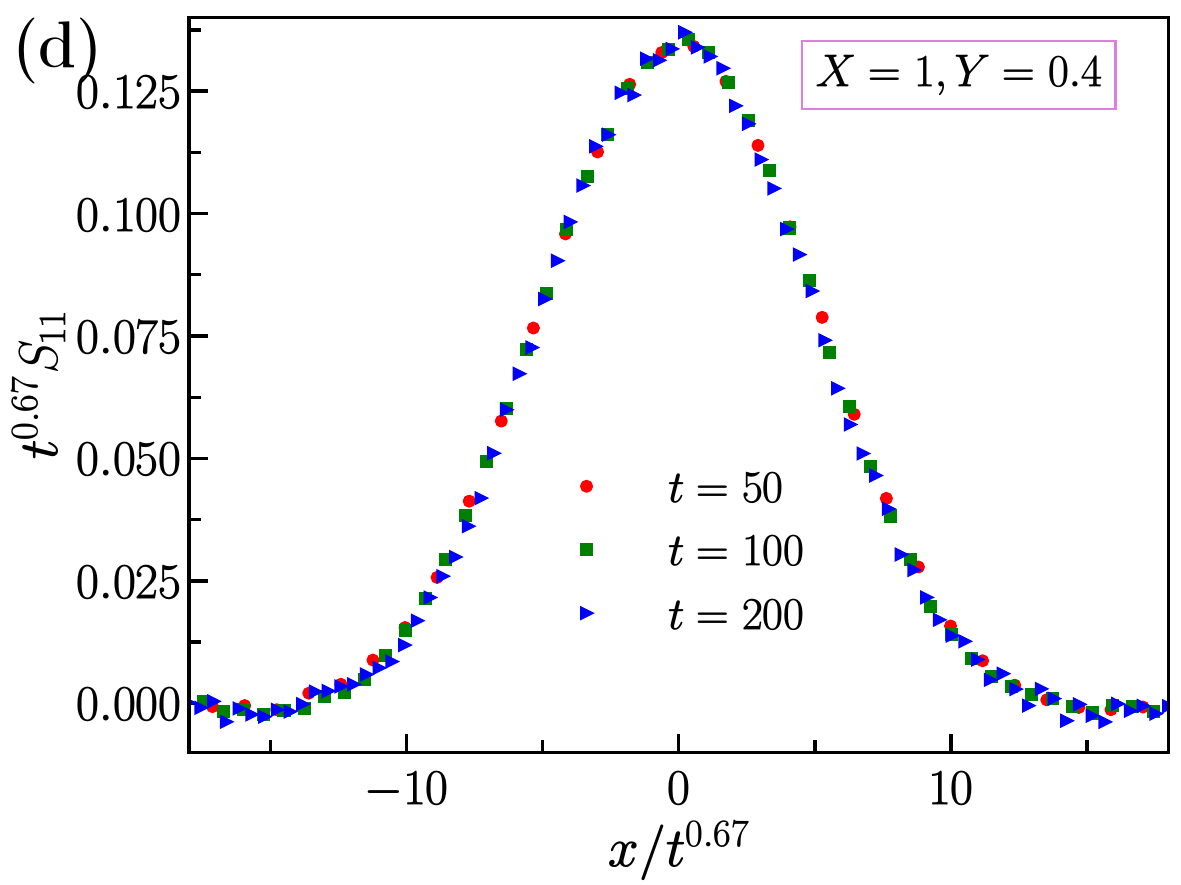}
		\end{subfigure}%
		\begin{subfigure}{0.33\linewidth}
			\includegraphics[width=\linewidth]{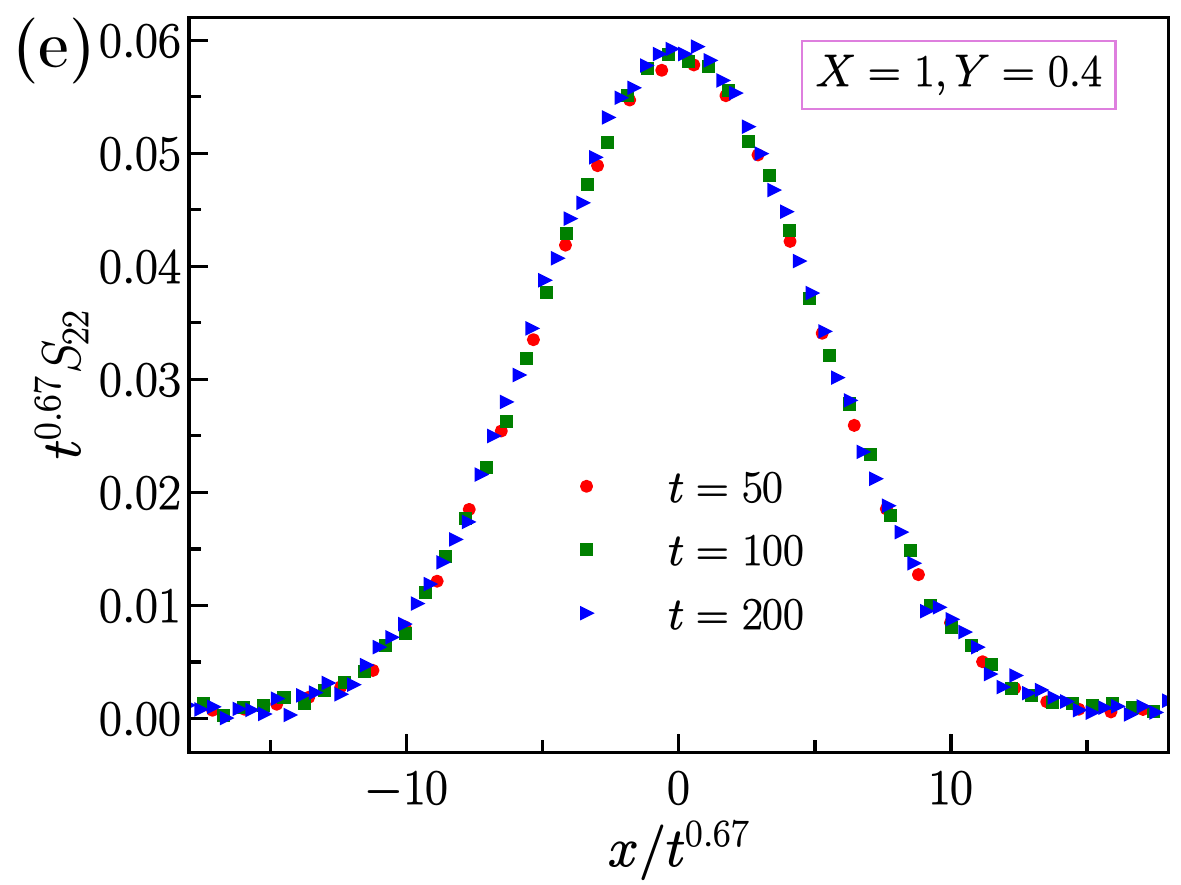}
		\end{subfigure}%
		\begin{subfigure}{0.33\linewidth}
			\includegraphics[width=\linewidth]{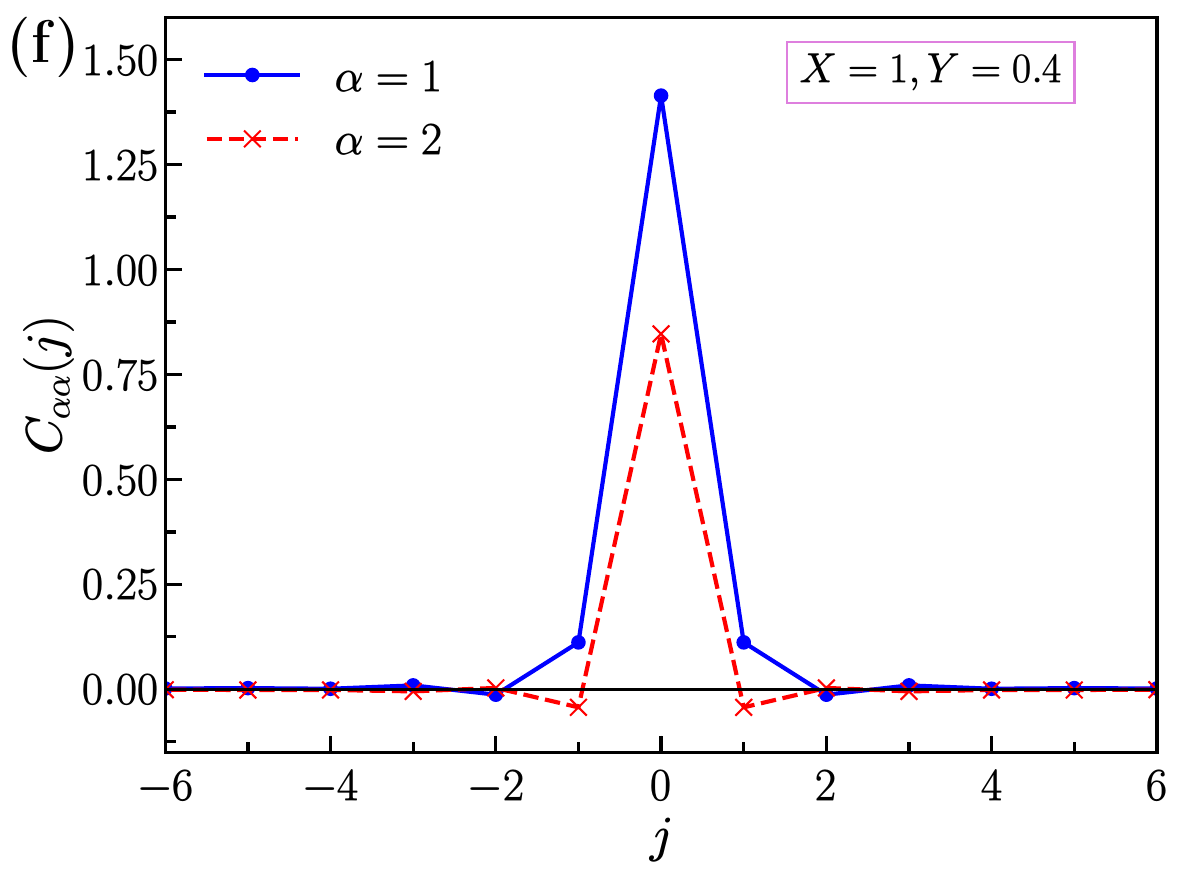}
		\end{subfigure}
		\begin{subfigure}{0.33\linewidth}
			\includegraphics[width=\linewidth]{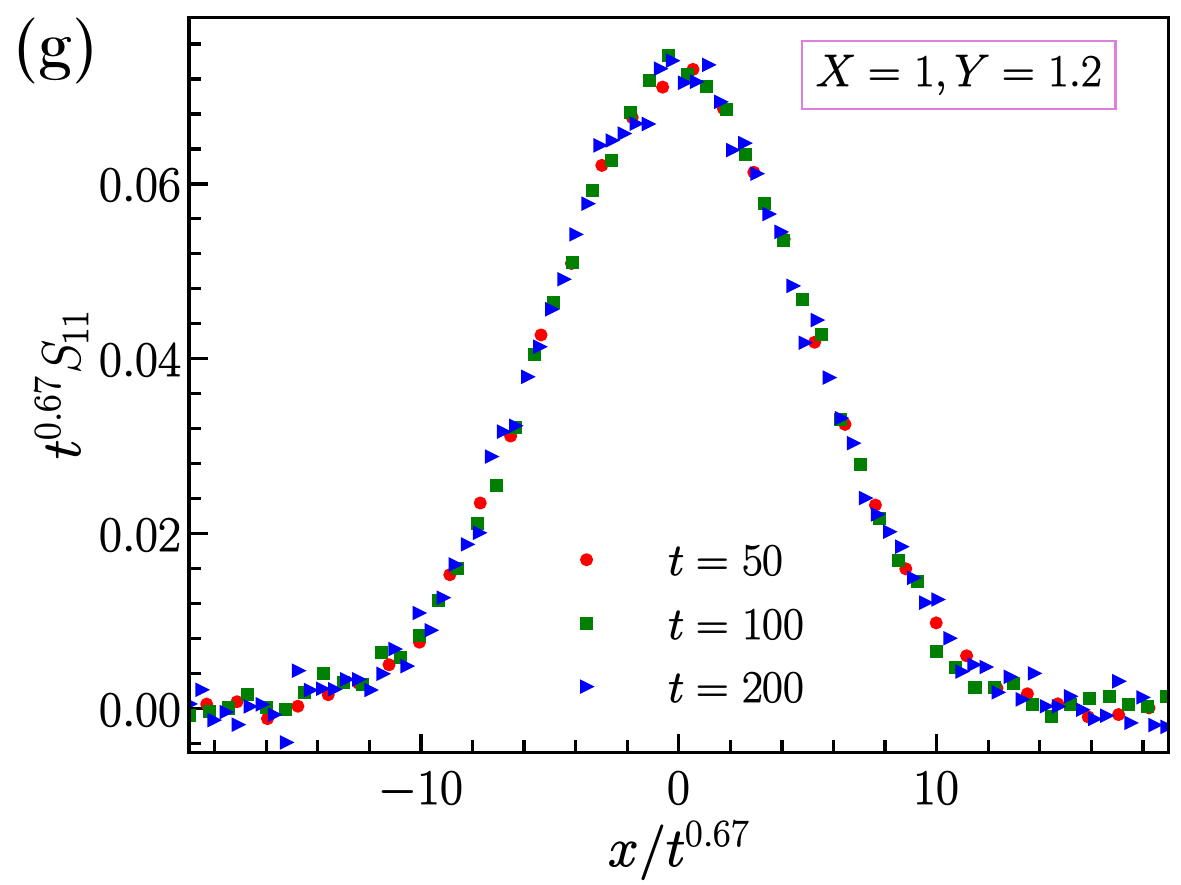}
		\end{subfigure}%
		\begin{subfigure}{0.33\linewidth}
			\includegraphics[width=\linewidth]{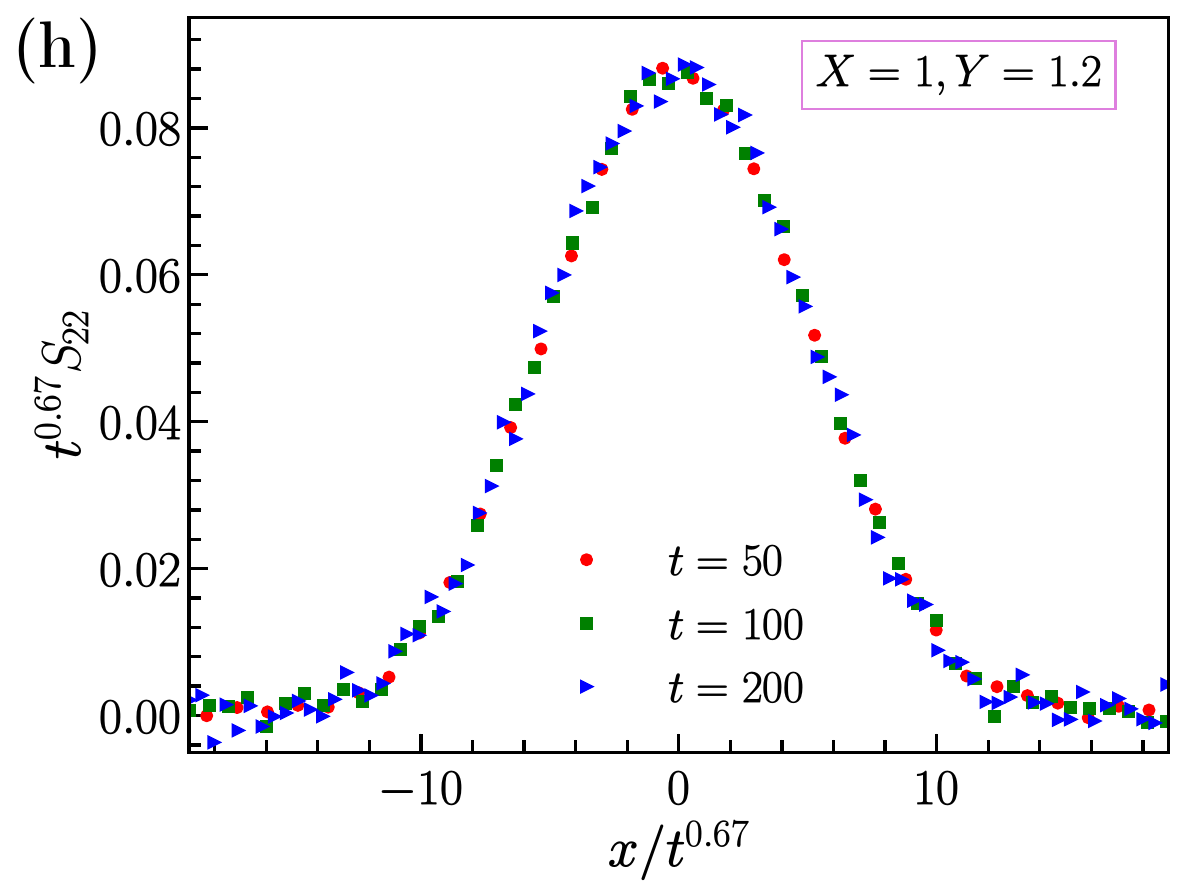}
		\end{subfigure}%
		\begin{subfigure}{0.33\linewidth}
			\includegraphics[width=\linewidth]{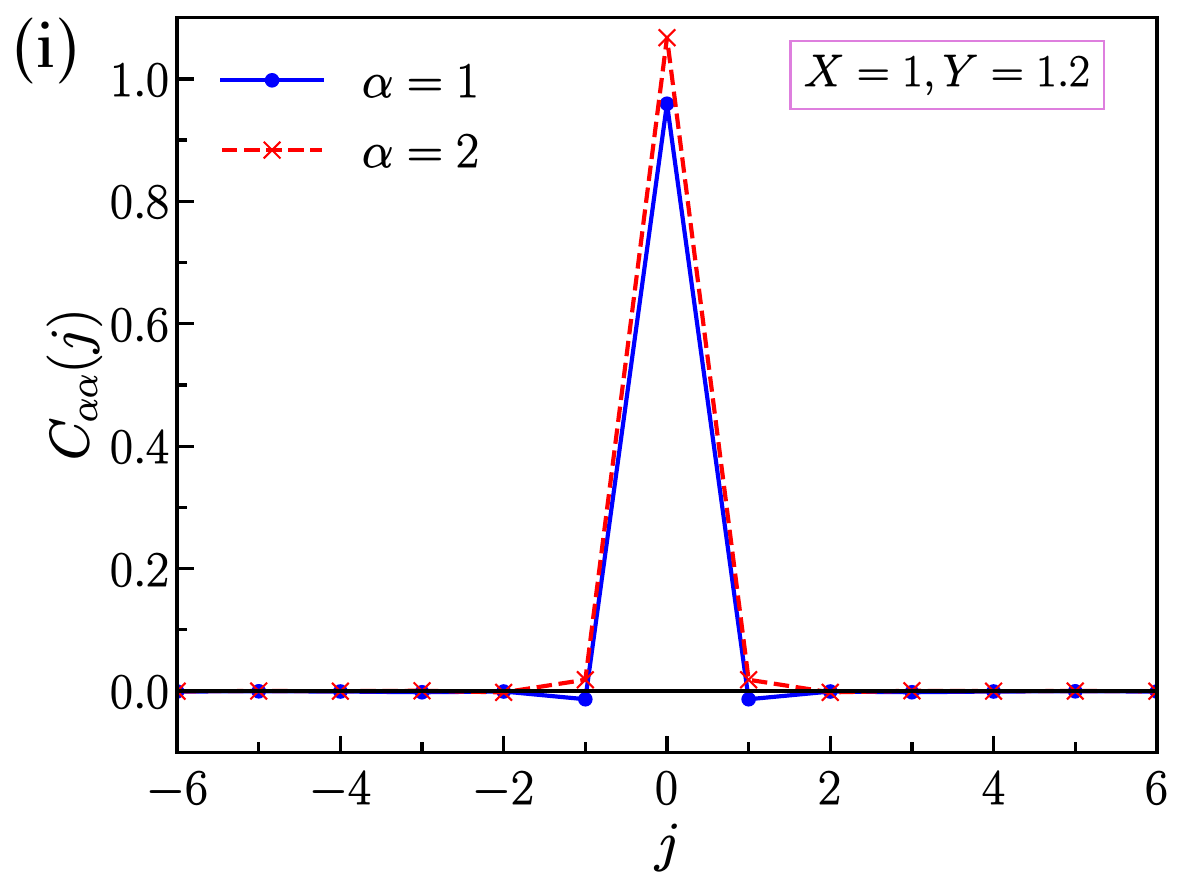}
		\end{subfigure}
        \caption{Simulations for parameters $X=1$, $Y= 1,0.4,1.2$.  We displayed   the steady state static correlators $C_{11}(j)$ and $C_{22}(j)$, (c), (f), (i). On the scale $t^{0.67}$ plotted are the spacetime correlations $S_{11}$, (a), (d), (g), and $S_{22}$, (b), (e), (h), at different times $t=50, 100, 200$. The parameter $\mathsfit{b}$ is provided in Table 1 and sampled are $10^4$ independent realizations.}
        \label{fig:cb1}
	\end{figure}
\end{center}

\begin{center}
	\begin{figure}[ht]
		\begin{subfigure}{0.33\linewidth}
			\includegraphics[width=\linewidth]{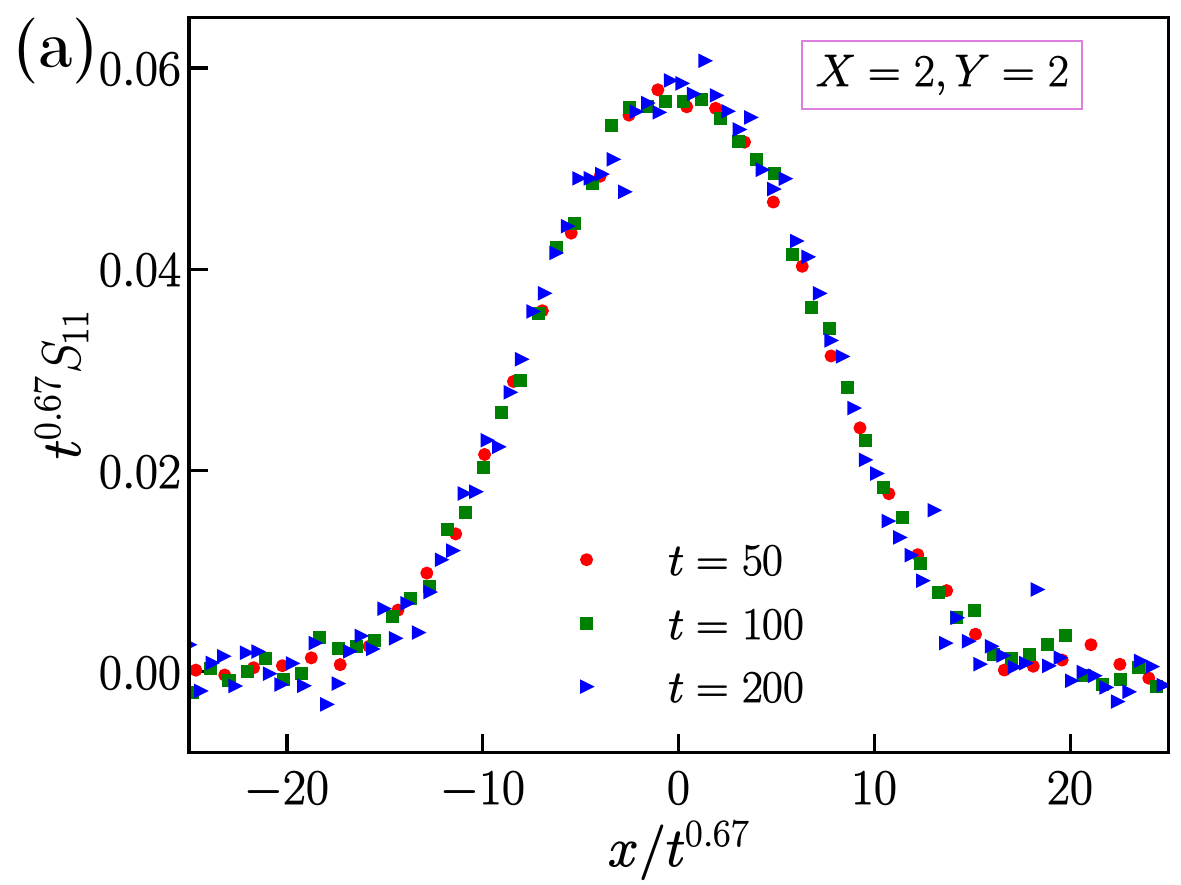}
		\end{subfigure}%
		\begin{subfigure}{0.33\linewidth}
			\includegraphics[width=\linewidth]{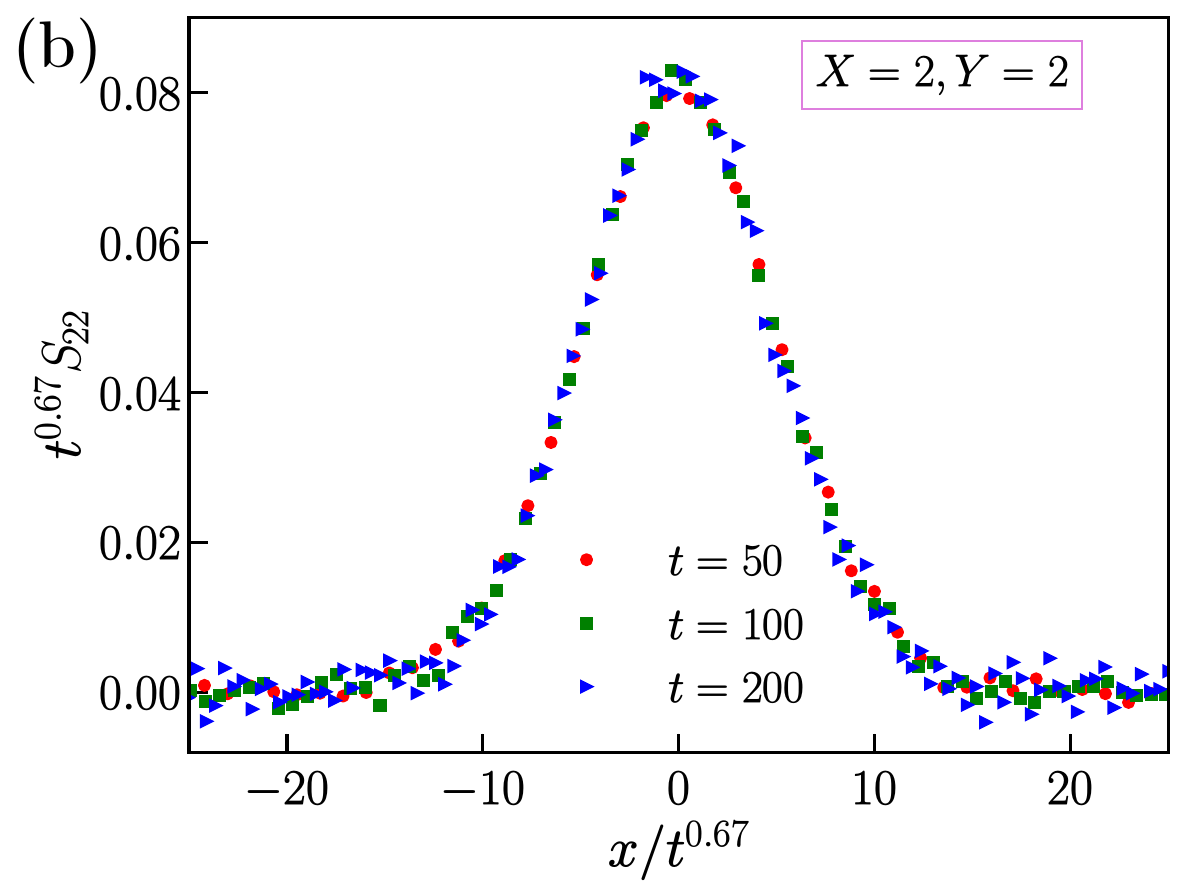}
		\end{subfigure}%
		\begin{subfigure}{0.33\linewidth}
			\includegraphics[width=\linewidth]{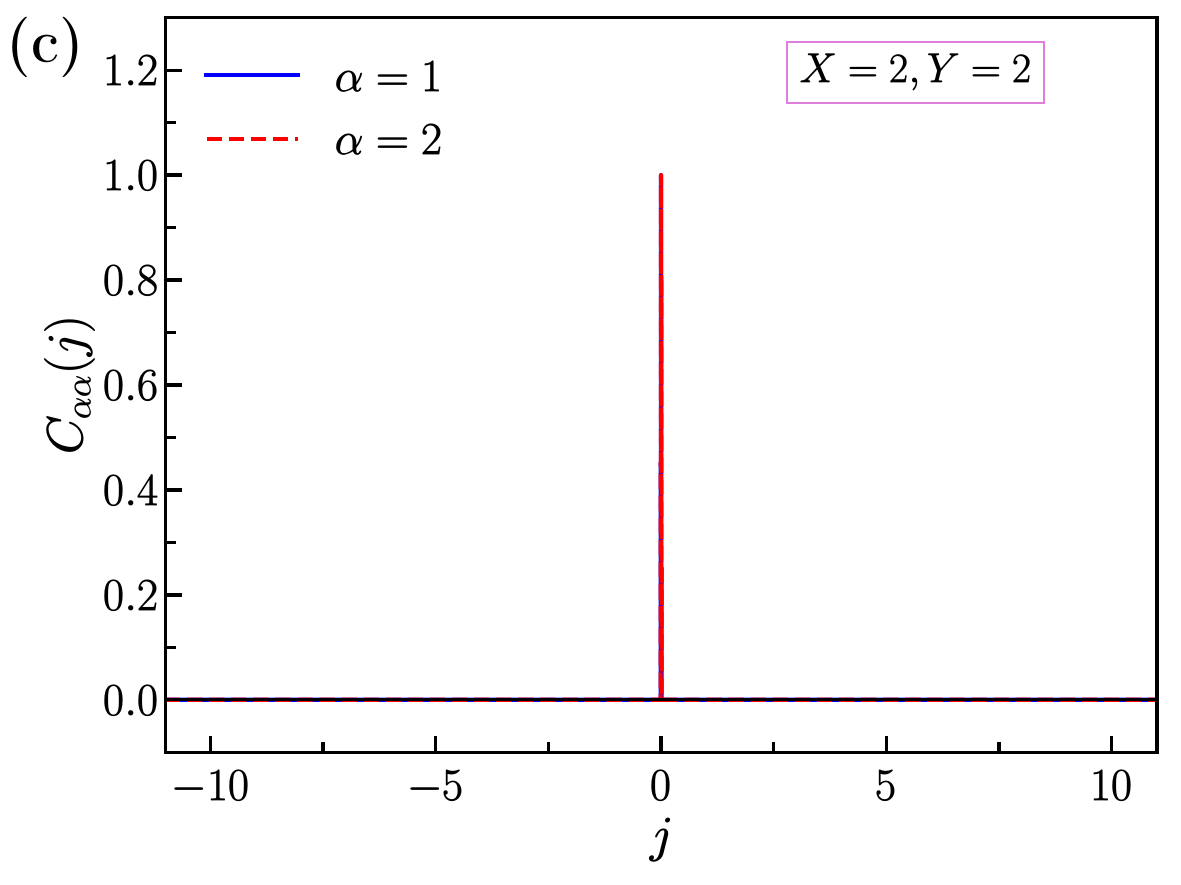}
		\end{subfigure}
            \begin{subfigure}{0.33\linewidth}
			\includegraphics[width=\linewidth]{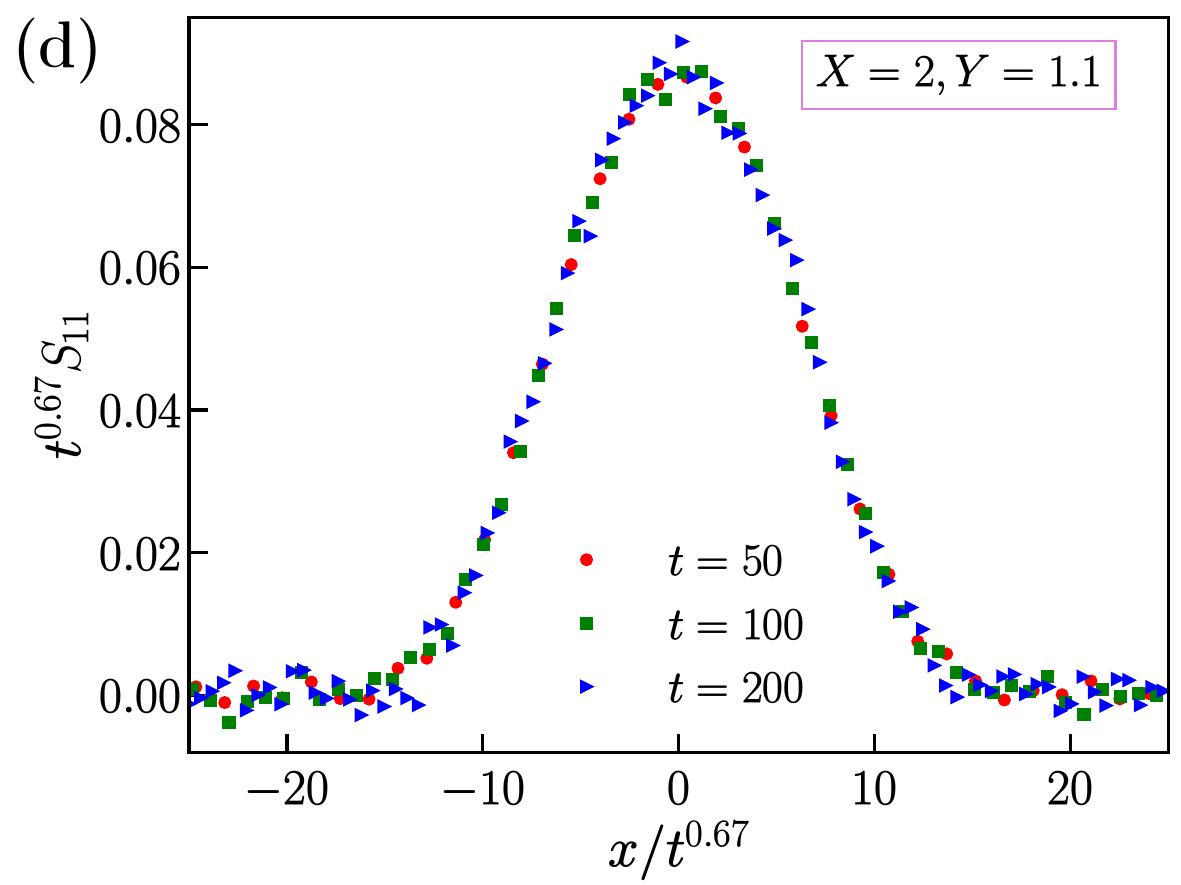}
		\end{subfigure}%
		\begin{subfigure}{0.33\linewidth}
			\includegraphics[width=\linewidth]{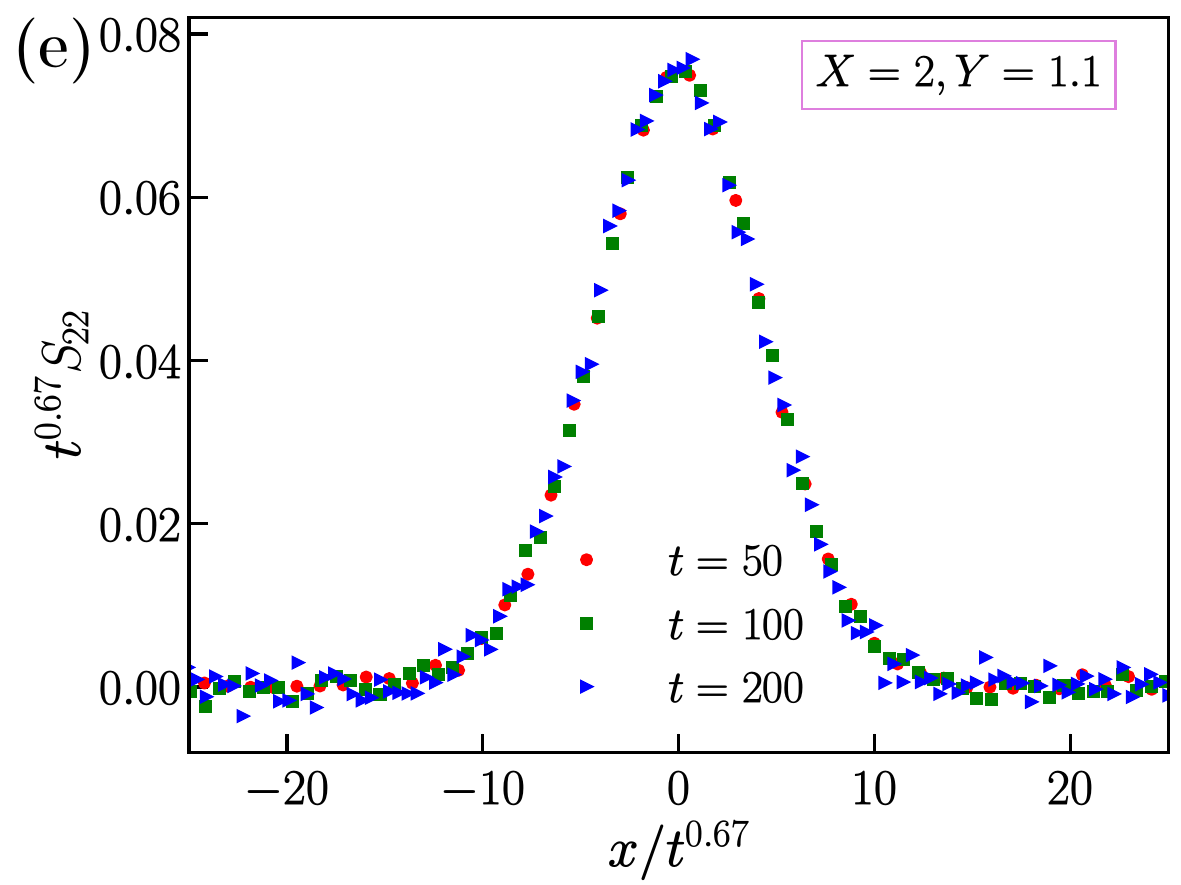}
		\end{subfigure}%
		\begin{subfigure}{0.33\linewidth}
			\includegraphics[width=\linewidth]{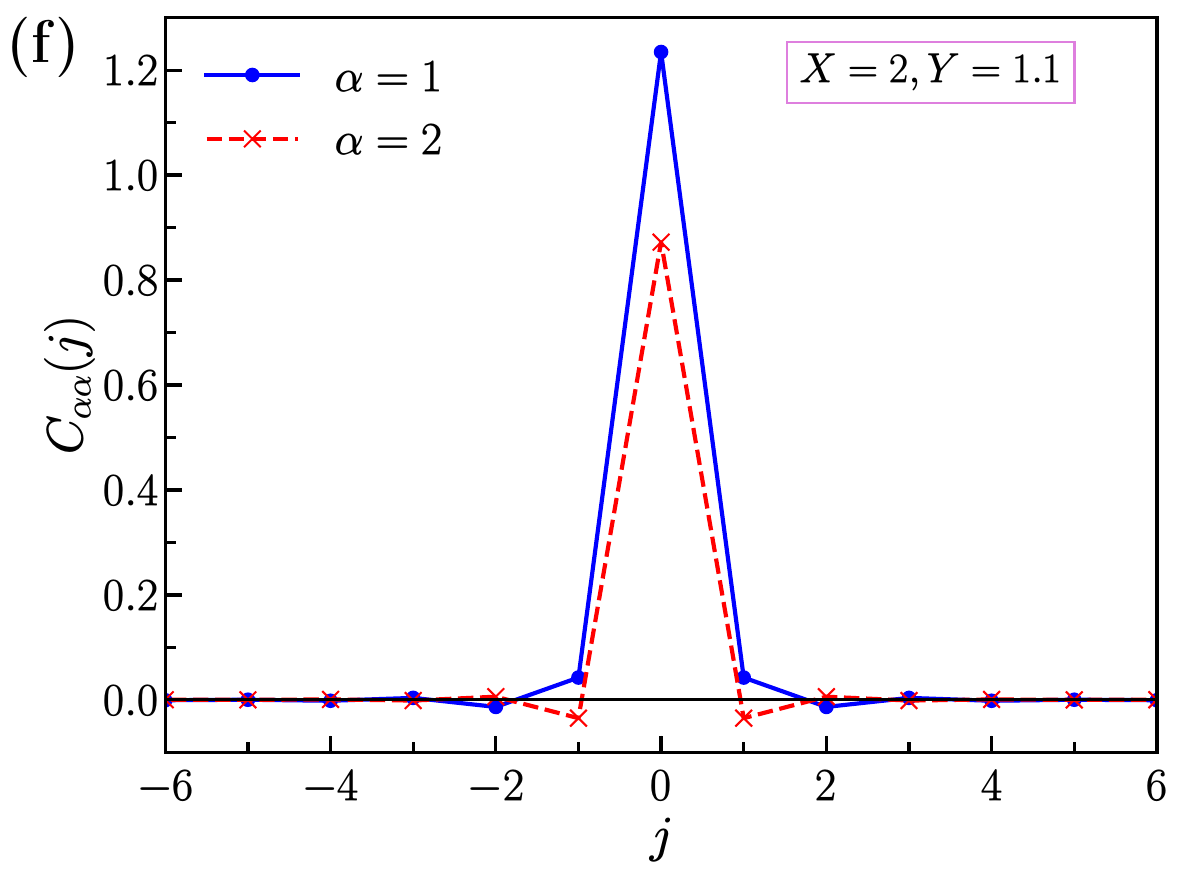}
		\end{subfigure}
            \begin{subfigure}{0.33\linewidth}
			\includegraphics[width=\linewidth]{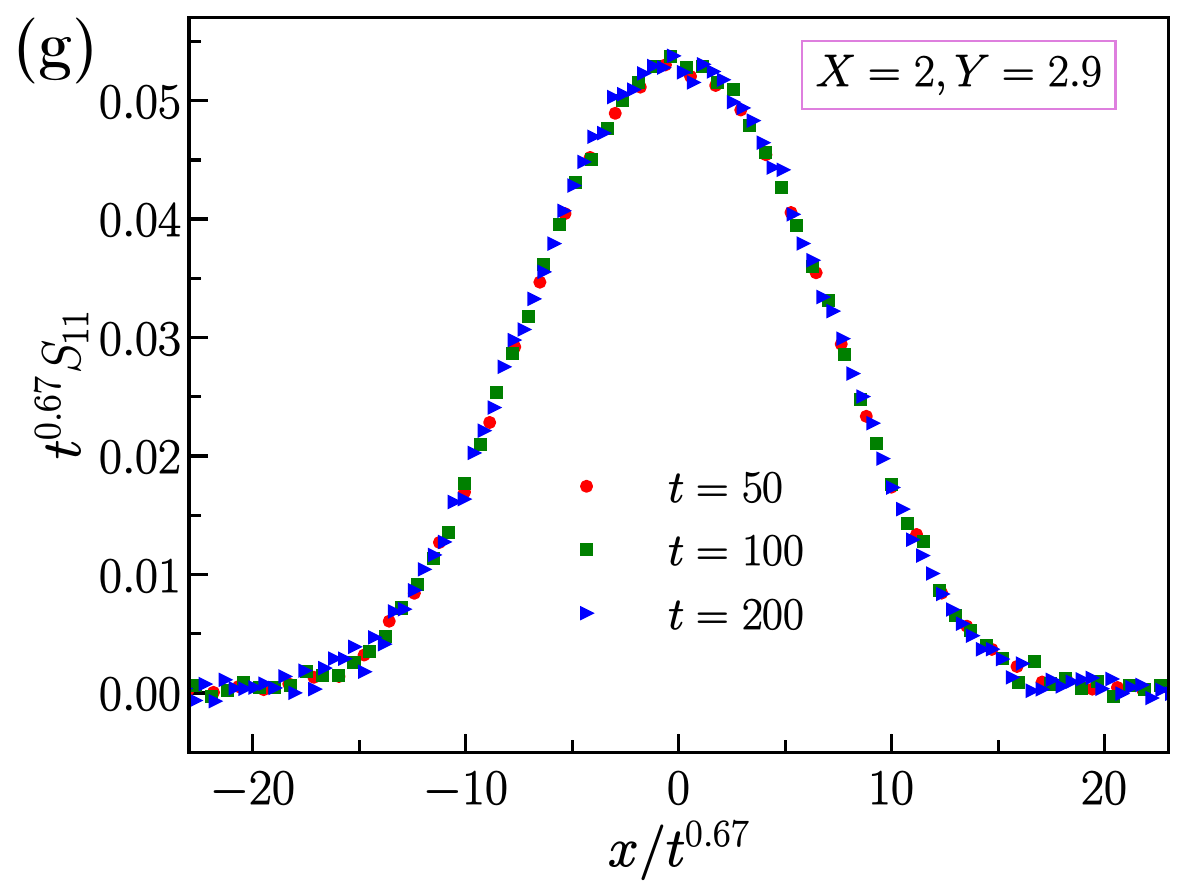}
		\end{subfigure}%
		\begin{subfigure}{0.33\linewidth}
			\includegraphics[width=\linewidth]{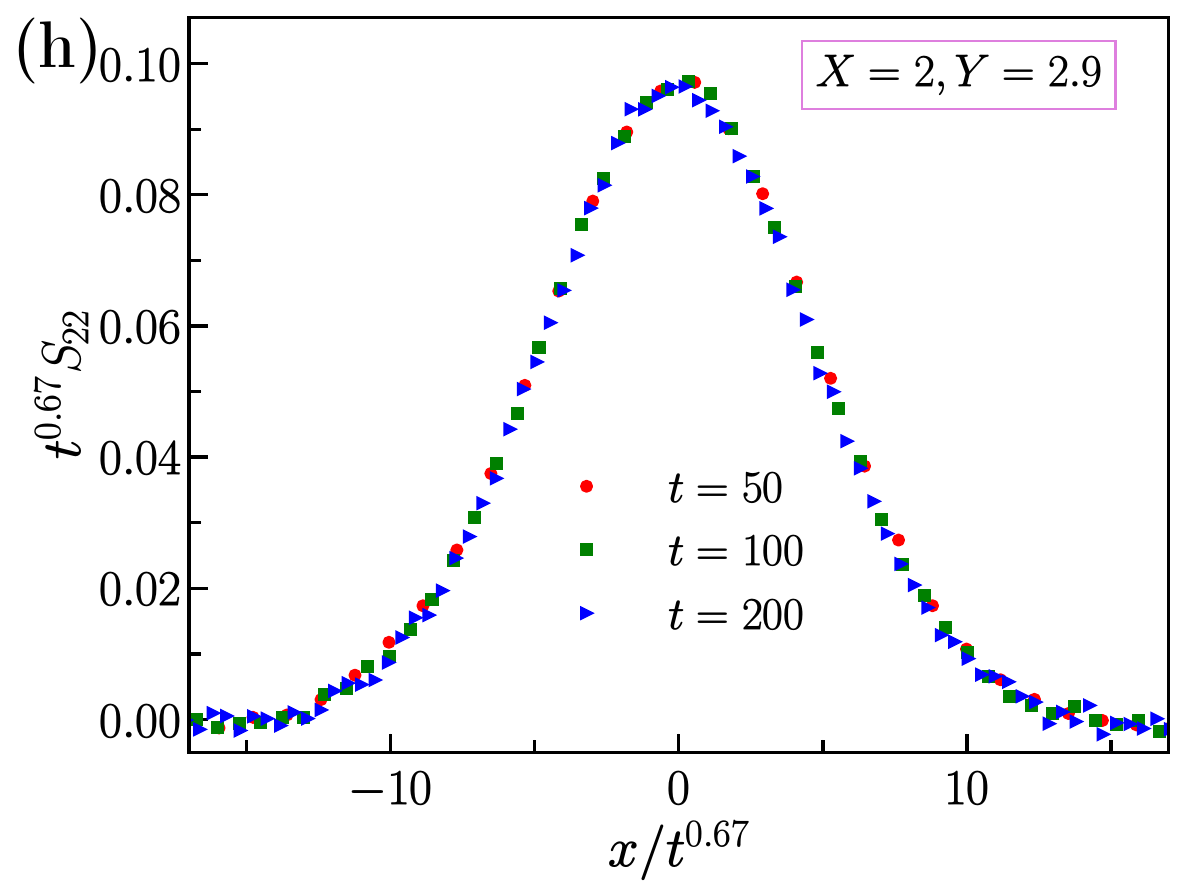}
		\end{subfigure}%
		\begin{subfigure}{0.33\linewidth}
			\includegraphics[width=\linewidth]{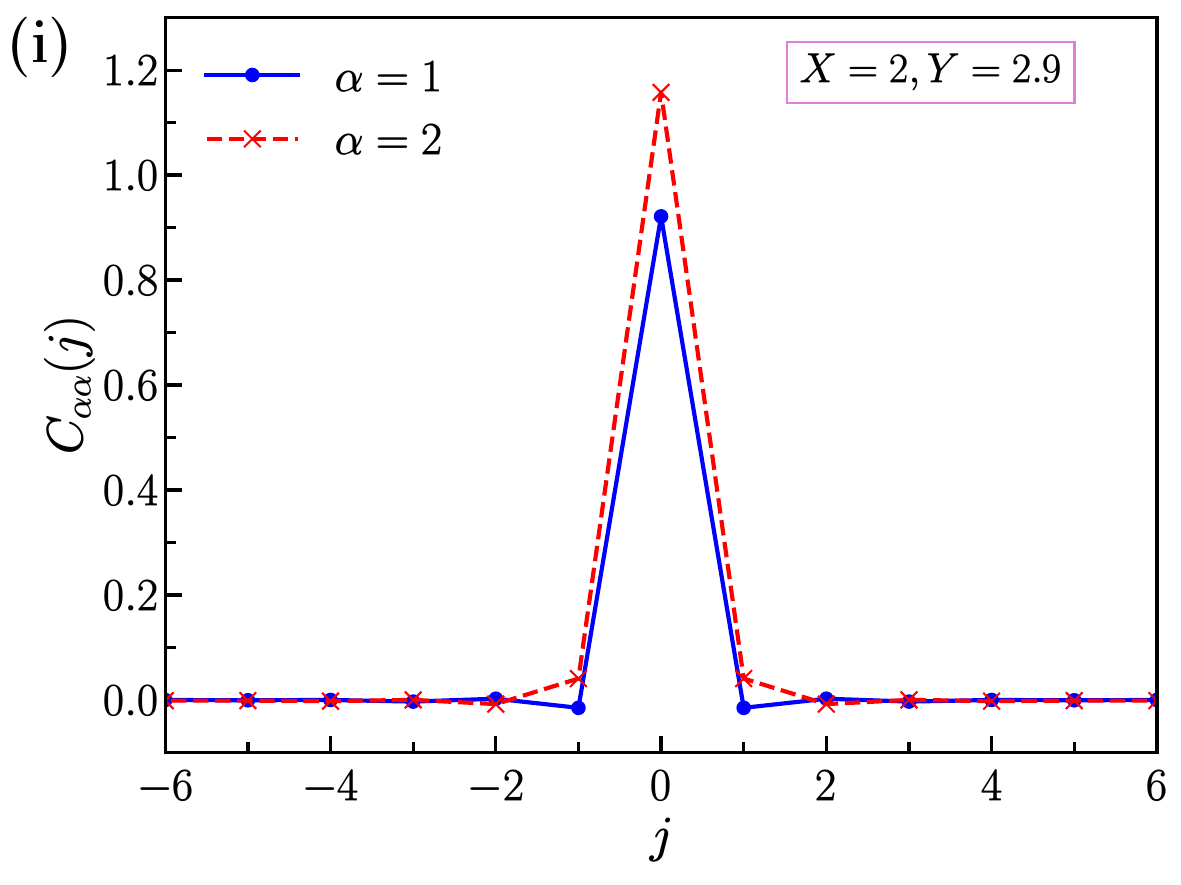}
		\end{subfigure}
		\caption{Simulations for parameters $X=2$, $Y= 2,1.1,2.9$.  We display  the steady state static correlators $C_{11}(j)$ and $C_{22}(j)$, (c), (f), (i). On the scale $t^{0.67}$ plotted are the spacetime correlations $S_{11}$, (a), (d), (g), and $S_{22}$, (b), (e), (h), at different times $t=50, 100, 200$. The parameter $\mathsfit{b}$ is provided in Table 1 and sampled are $10^4$ independent realizations. }
        \label{fig:cb2}
	\end{figure}
\end{center}

According to the scheme outlined above, simulated are $C_{\alpha\alpha}(j,t_\mathrm{eq})$ and $S_{\alpha\alpha}(j,t)$ for the six parameter points from the Table.  Within numerical error bars, the dynamical correlator is scaled with the exponent  $\tfrac{3}{2}$ and thereby provides information on the scaling functions $g_{\alpha}$.
For the cyclic parameters along the diagonal, the susceptibilities are $c_1 =1$, $c_2=1 $. As an important control check, we confirm that $C_{\alpha\alpha}(j,t)$ is concentrated at the single site $j= 0$ of the numerical grid. For non-cyclic
parameters, $C_{\alpha\alpha}(j,t_\mathrm{eq})$ is broadened and its susceptibility $c_\alpha$ is determined by Eq. \eqref{3.6}. The results are displayed in Figs. \ref{fig:cb1} and \ref{fig:cb2}. 

\begin{table}[h]
\begin{center}
\renewcommand*{\arraystretch}{1.4}
\begin{tabular}{|c|c|c|c|c|c|c|c|c|}
\hline
	Fig.  &   $\mathsfit{b}$ &  $X$ &  $Y$ & Symbol  & $c_1$  & $c_2$  & $c_2p_1 / c_1 p_2$ & $c_2 X/c_1 Y$\\
\hline
	$3$ & $4$ & $1$ & $1$ & red lozenge & $1.0$ & $1.0$ & $1.0$   & 1.00\\
\hline
    $3$ & $5$ & $1$& $0.4$& magenta asterisk & $1.69$ & $0.72$ & $0.97$  & $1.07$     \\
    \hline
	$3$ & $4$ &  $1$ & $1.2$ & pink asterisk & $0.91$ & $1.10$ & $1.00$ & 1.01\\ 
	\hline\hline
	$4$ & $3$ &  $2$ & $2$ & blue lozenge & $1.0$ & $1.0$ & $0.75$ & 1.00\\
	\hline
	$4$ & $3$ &  $2$ & $1.1$ & blue asterisk & $1.30$ & $0.82$ & $0.73$ & 1.15\\ 
    \hline
        $4$ & $2.5$ &  $2$ & $2.9$ & purple asterisk & $0.88$ & $1.20$ & $0.75$  & $0.94$   \\
	\hline
    \end{tabular}\bigskip\bigskip
    \caption{Ratios $c_2 p_1 / c_1 p_2$ and $c_2 X / c_1 Y$ at parameters used in numerical simulations of \eqref{3.1}. Here $c_\alpha$ is the static susceptibility and $p_\alpha$ the height at the origin of the peak $\alpha$. There are two universality classes, $X=1$, $X=2$, and the symbols correspond to the ones displayed in Fig.~\ref{fig1}. The values of $c_\alpha$ are computed using a quadratic fit for $\hat{C}_{\alpha\alpha}(k,t_\mathrm{eq})$, compare with \eqref{3.6} and explanations below. The ratios $c_2 p_1 / c_1 p_2$ and $c_2 X / c_1 Y$ are observed to be in reasonable good agreement in either class.}
\end{center}
\end{table}
\clearpage

One first notes that the static correlator, while broadened, decays very quickly to zero, thereby confirming a key assumption of our theory. Secondly within error bars the scaling exponent of $0.67$ is well confirmed throughout. At $(1,1)$ the peaks have the same height and become asymmetric when moving to $(1,0.4)$ and $(1,1.2)$. This is a nonuniversal feature which is compensated by $c_1,c_2 \neq 1$. At $(2,2)$ the fixed point has already an intrinsic asymmetry, which is reduced by moving to  $(2,1.1)$ and  enforced at $(1,2.9)$,
again nonuniversal features. 
     
\clearpage

From our analysis we concluded that the ratios $c_2p_1 / c_1 p_2$ and $c_2 X/c_1 Y$ should be independent of $Y$ for given $X$. In the Table the measured values of $c_\alpha$ are listed and from Figures \ref{fig:cb1} and \ref{fig:cb2} one reads off the value of $p_\alpha$. The resulting ratios 
are in reasonable good agreement with theory. However, based on prior experience,   
a much stronger criterion is to compare the full scaling functions, the results of which will be reported next.

\subsection{Comparison of scaling functions at ${X=1,2}$}
\label{sec3.2}
\begin{center}
	\begin{figure}[ht]
		\begin{subfigure}{0.5\linewidth}
			\includegraphics[width=\linewidth]{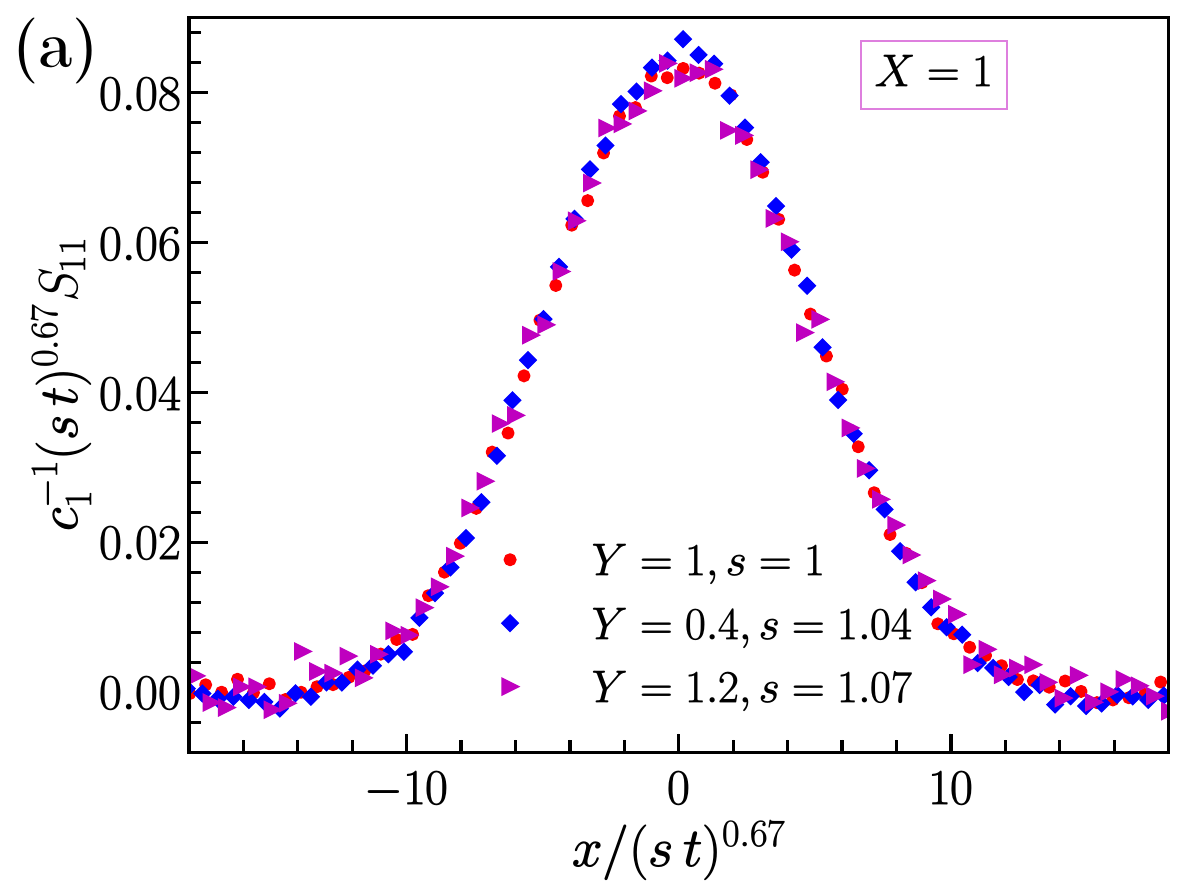}
		\end{subfigure}%
		\begin{subfigure}{0.5\linewidth}
			\includegraphics[width=\linewidth]{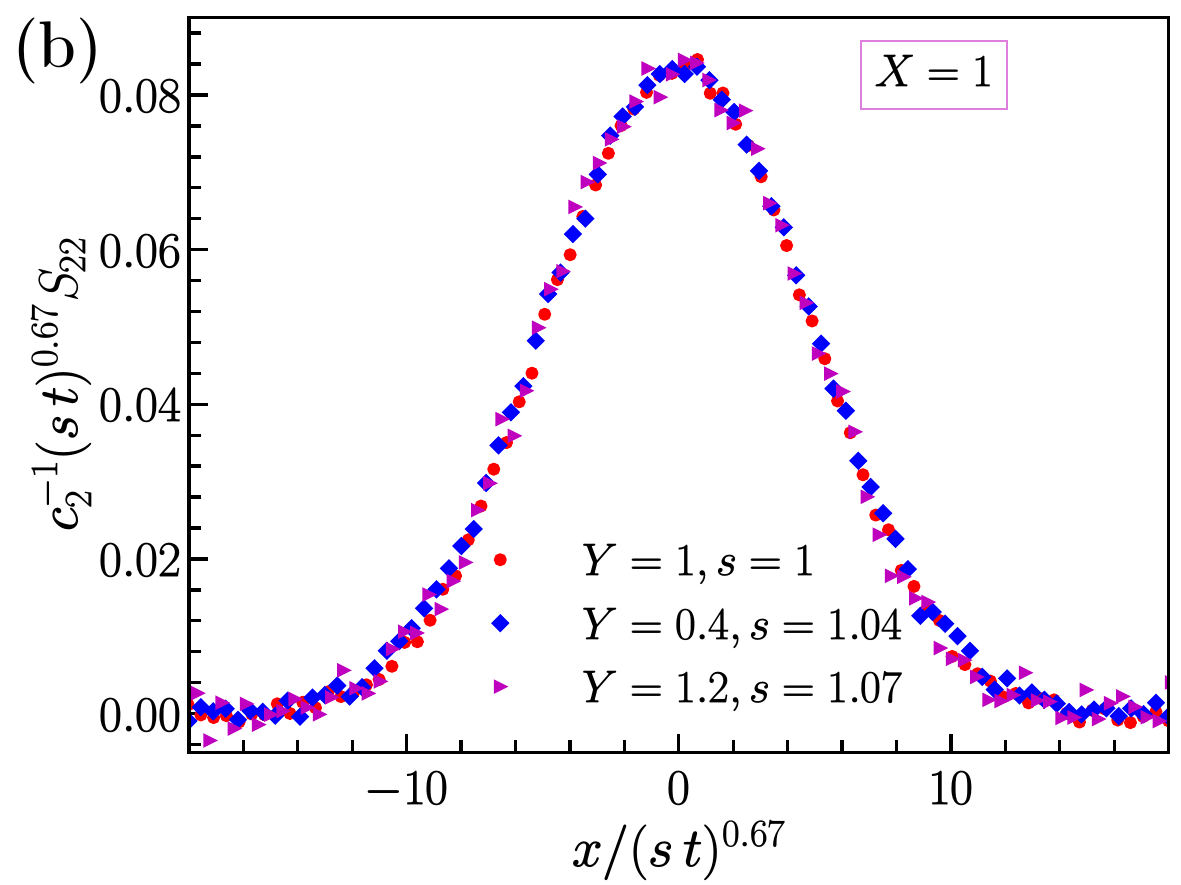}
		\end{subfigure}
		\caption{Universal scaling function for $X=1$, $Y = 1,0.4,1.2$. For the longest available time $t = 200$, the six curves from Figure \ref{fig:cb1} are multiplied by $1/c_\alpha$ and the free time scale $s$ is optimized.  } 
        \label{fig:eb1}
	\end{figure}
\end{center}

\begin{center}
	\begin{figure}[ht]
		\begin{subfigure}{0.5\linewidth}
			\includegraphics[width=\linewidth]{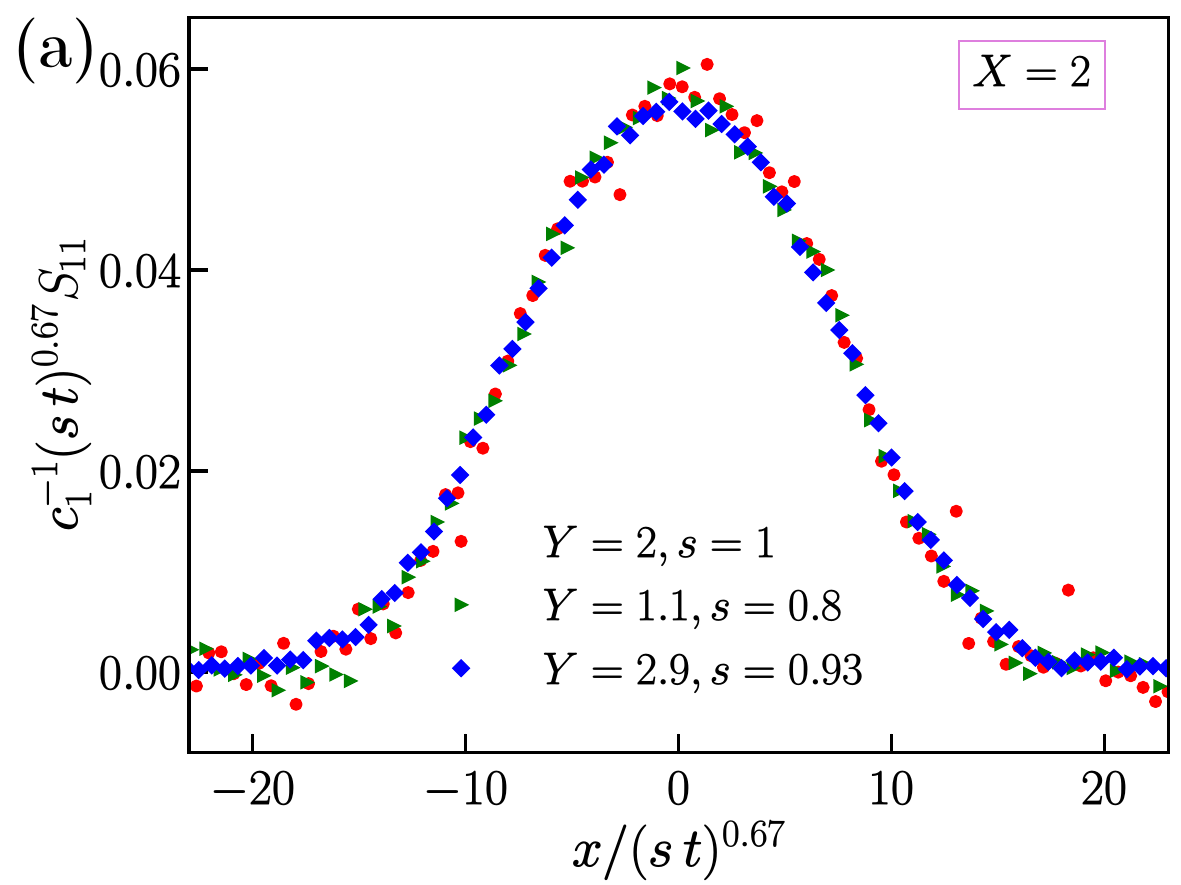}
		\end{subfigure}%
		\begin{subfigure}{0.5\linewidth}
			\includegraphics[width=\linewidth]{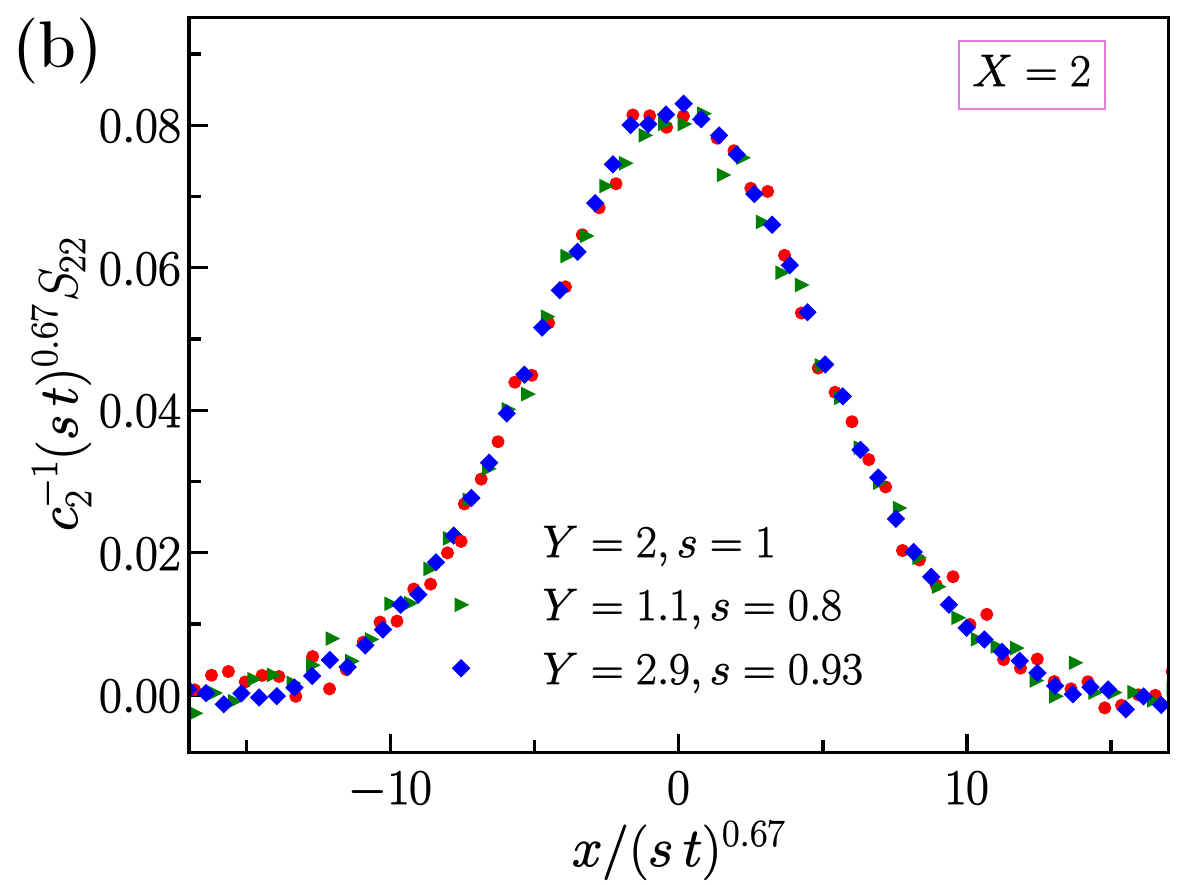}
		\end{subfigure}
		\caption{Universal scaling function for $X=2$, $Y = 1,1.1, 2.9$. For the longest available time $t = 200$, the six curves from Figure \ref{fig:cb2} are multiplied by $1/c_\alpha$ and the free time scale $s$ is optimized.} 
        \label{fig:eb2}
	\end{figure}
\end{center}

For the KPZ equation there is a well understood scaling theory,
which predicts the nonuniversal model-dependent coefficients. In numerical simulations, in particularly away from integrability \cite{schmidt}, it is common experience that the scaling function has already the predicted shape, while the nonuniversal coefficients show still considerable deviations. Therefore when fitting theory with numerics one introduces a time scale as single free parameter.  Ideally, for very long times this factor should be $1$. We adopt the same procedure for our two-component system and introduce the free parameter $s$ through the scaled time $st$ and optimize so to have maximal agreement. Since there is only one time, the scale parameter has to be the same for both components. 
Separately for $X=1$ and $X=2$, the  dynamical correlators from Figures \ref{fig:cb1}, \ref{fig:cb2} are normalized by $1/c_{\alpha}$ and the parameter $s$ is introduced. In Figures \ref{fig:eb1}, \ref{fig:eb2} the optimal fit is shown and the parameter $s$ is recorded.
\clearpage

Within numerical errors the fit is viewed as convincing. In \cite{RDK24}, we reported on a similar comparison between the scaling functions of a two lane-lattice gas and the corresponding 
two-component KPZ equation. Both models are cyclic and no equilibration step is required. On the other hand, beyond the factor $1/c_\alpha$, a rotation by $\pi/4$
had to be implemented. In this case, the coincidence of the two scaling functions is more striking 
than the one of Figures \ref{fig:eb1}, \ref{fig:eb2}. Presumably, further averaging would improve the agreement. 
\subsection{Beyond the subspace $\{T=1\}$}
\label{sec3.3}
To explore $T\neq 1$, we follow the same protocol used when varying $Y$. Direct simulations are carried out for the parameters $(2,2,T)$ with $T = 0.5,1,2$. Since the parameters are cyclic the equilibration time $t_\mathrm{eq}= 0$. To account for the possible non-universal factors  when comparing with $T=1$, the scale $s$ is introduced as a free parameter. For the two values $T= 0.5,2$  
a good fit with the scaling function at $(2,2,1)$ is achieved. To be noted, the scale factor is $s= 0.65$ for $T=0.5$  and $s= 0.6$ for $T=2$.
More simulations are in demand. But at least for the chosen parameter values, a physically natural scenario is confirmed: For the long time behavior 
locally the strength of the diffusion term and of the noise term enter only through the thereby determined effective susceptibility.

\begin{figure}[ht]
    \begin{center}
            \begin{subfigure}{0.45\linewidth}
			\includegraphics[width=\linewidth]{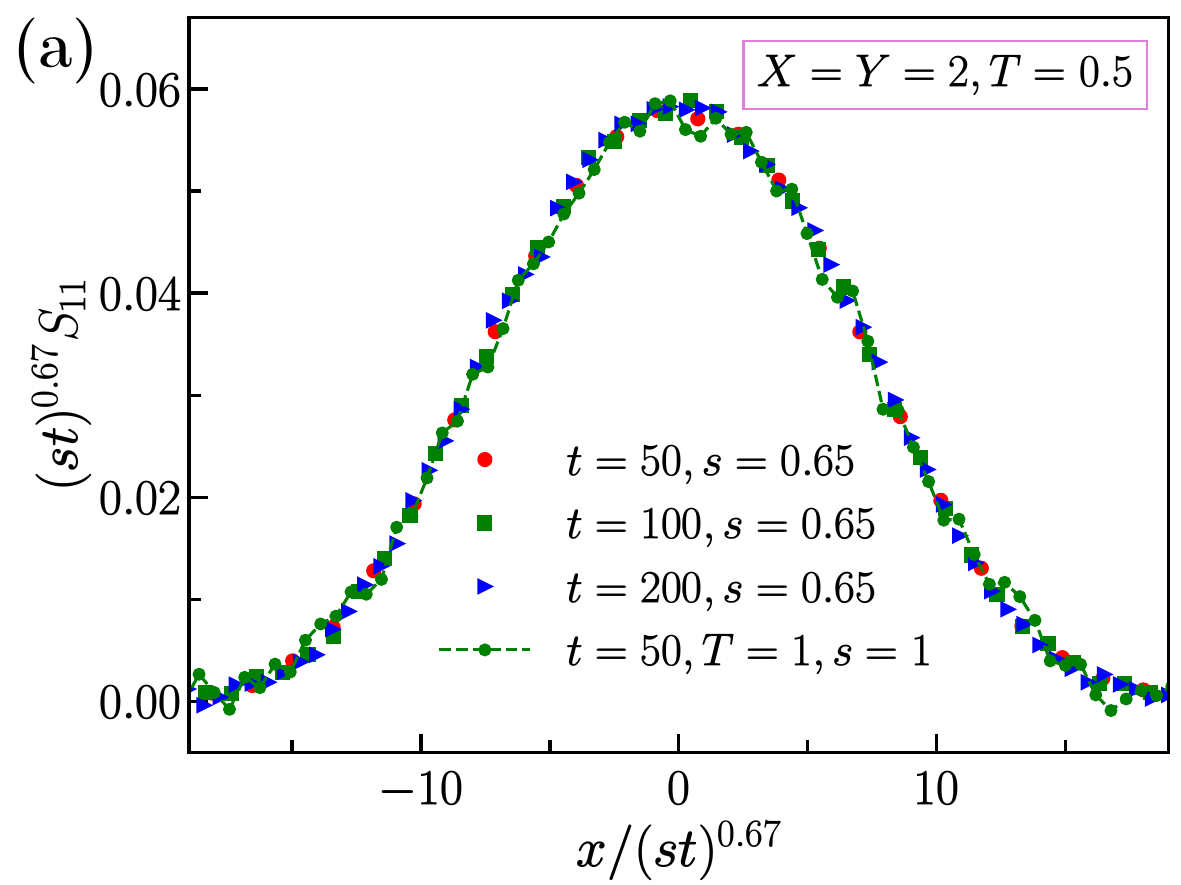}
		\end{subfigure}%
            \begin{subfigure}{0.45\linewidth}
			\includegraphics[width=\linewidth]{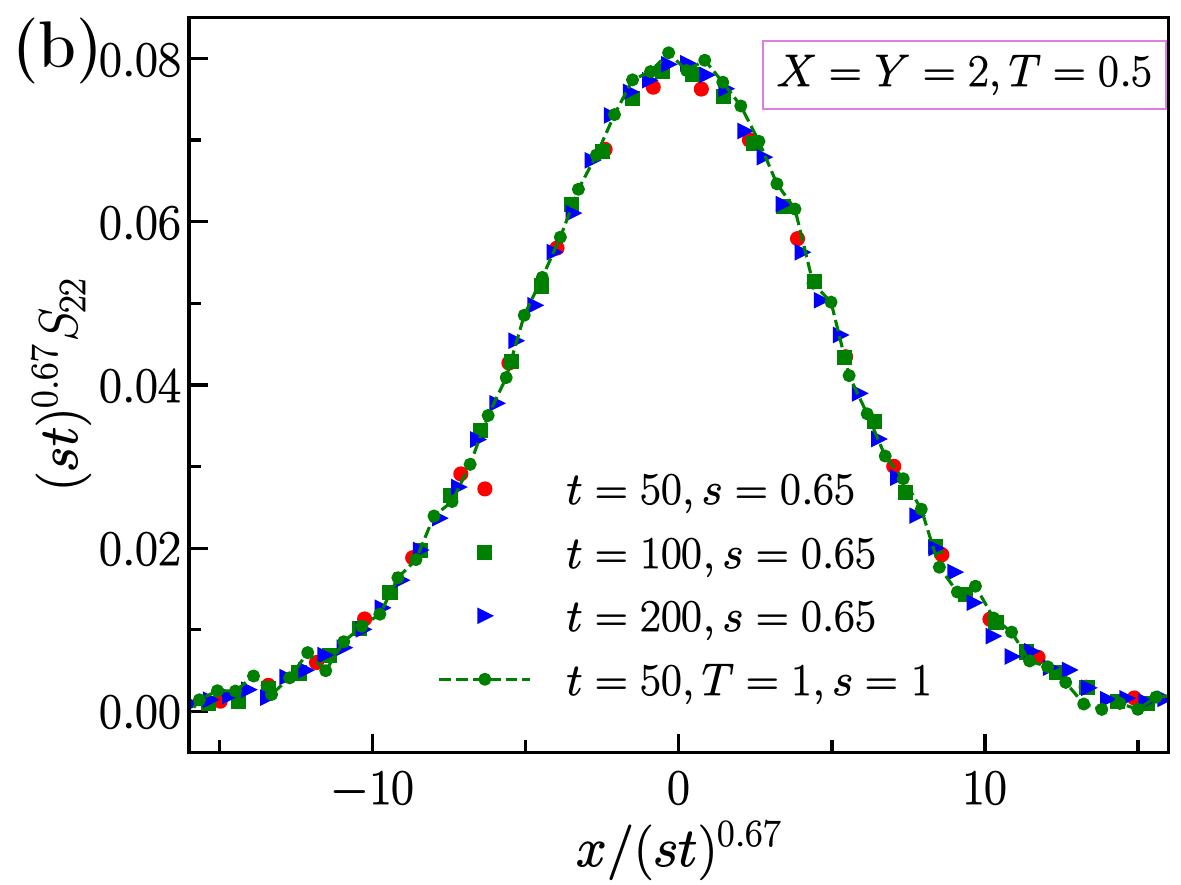}
		\end{subfigure}
		\begin{subfigure}{0.45\linewidth}
			\includegraphics[width=\linewidth]{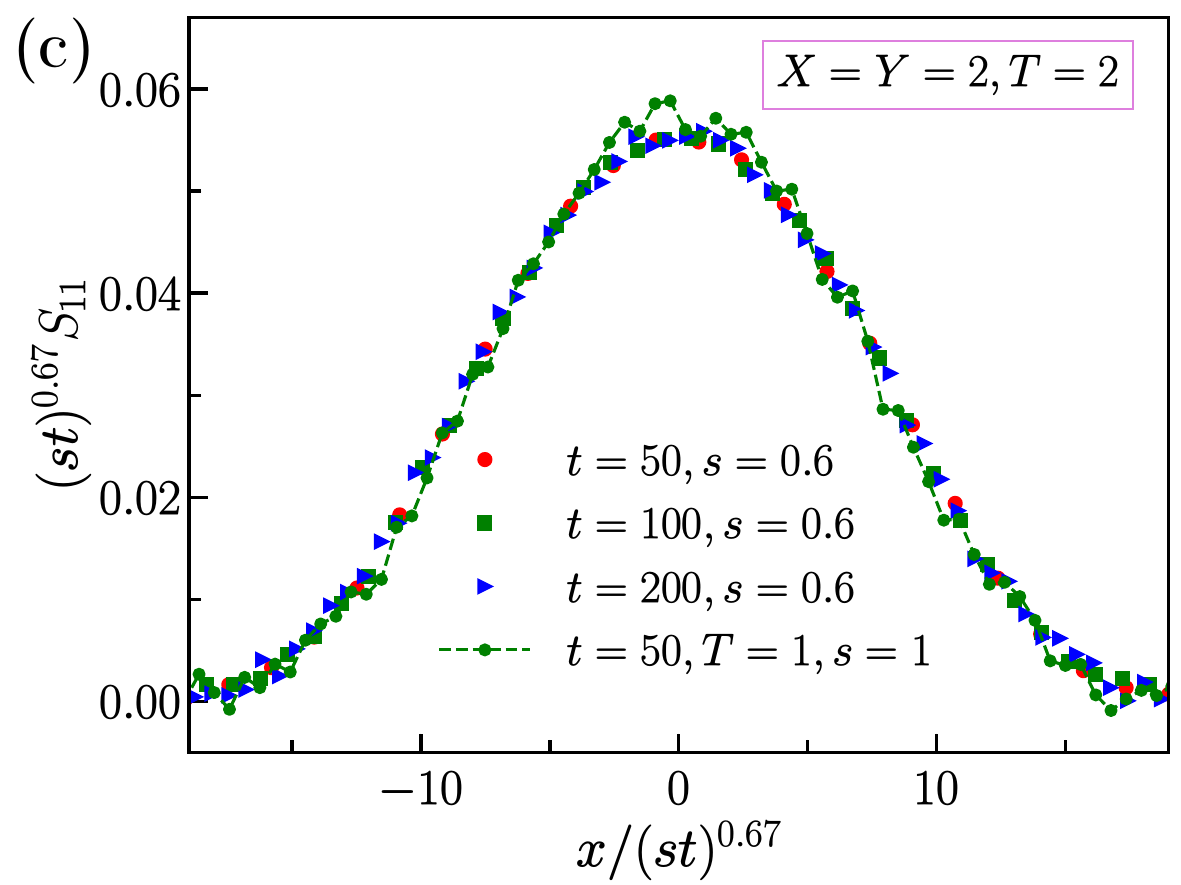}
		\end{subfigure}%
		\begin{subfigure}{0.45\linewidth}
			\includegraphics[width=\linewidth]{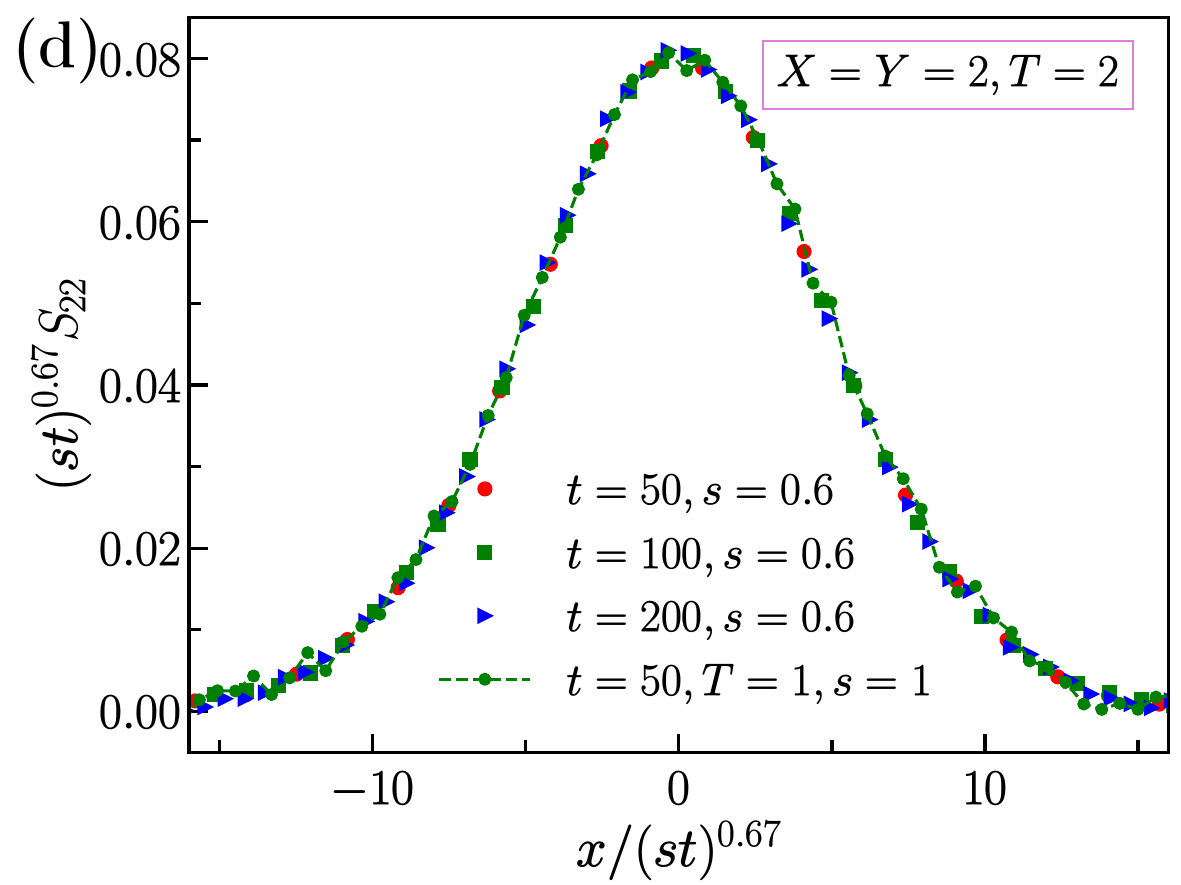}
		\end{subfigure}
        \caption{Simulations for parameters $X=2$, $Y= 2$ with $T=2,0.5$. On the scale $t^{0.67}$ plotted are the spacetime correlations $S_{11}$, (a), (c), and $S_{22}$, (b), (d), at different times $t=50, 100, 200$. The parameter $\mathsfit{b} = 2$ and sampled are $5\times10^4$ independent realizations.
    The data are compared with the fixed point parameters $(2,2,1)$ as displayed in top of Fig. \ref{fig:cb2}.}
        \label{fig:cb5}
        \end{center}
\end{figure}

\section{Discussions}
\label{sec4}
\setcounter{equation}{0} 
We return to the coupled KPZ equations 
\begin{equation}\label{4.1}
    \begin{aligned}
	\partial_t h_1 &=    2 X  (\partial_x h_1)(\partial_x h_2) + \tfrac{1}{2} T\partial_x^2 h_1+ \sqrt{T}\xi_1 , \\
	\partial_t h_2 &=  Y  (\partial_xh_1)^2 + (\partial_xh_2)^2 +  \tfrac{1}{2}\partial_x^2 h_2  +\xi_2.
    \end{aligned}
\end{equation} 
Summarizing our main result, we argued that the universality classes are labeled by the parameters $(X,X,1)$. Because of cyclicity, for both components $\chi = \tfrac{1}{2}$.
On the other hand, for the dynamical exponent and the scaling functions we have to rely on direct numerical simulations. Within the available numerical precision, the dynamical exponent  $z = \tfrac{3}{2}$ is used when comparing our data with theoretical results.
 
 D.~Erta\c{s} and M.~Kardar studied the same model and worked out the RG flow in one-loop approximation. They obtained flow equations for the 7 parameters displayed in Eq. \eqref{A.5}. As part of the discussions we explain their results in more detail.
 In the recent posting \cite{weinberger2024}, H.~Weinberger et al. reworked the flow equation, observing that in fact only  four coefficients are needed, namely $X,Y,T$ and an additional global factor, denoted by $Z$, in front of the nonlinearity. 
As shown in \ref{A}, the number of dimensionless parameters can be even further reduced to three, namely $(X,Y,T)$. Measuring the RG coarse-graining length $\ell$ in units of the length $1/(4 \pi \ell)=\sigma_2^2 \Gamma_2^2/(4 \pi D_2^3)$, see \ref{A}, and in agreement with \cite{1993-ertas-kardar}, \cite{weinberger2024},  the following flow equations hold: 
\begin{eqnarray}\label{4.2}
&&\frac{d}{d\ell}T= \frac{-2X((T-3)X-3T Y +Y)}{(1+T)^2} -T -X Y,\nonumber\\
&&\frac{d}{d\ell}X =-\frac{4(X-1)X(X-Y)}{(1+T)^2},\nonumber\\
&&\frac{d}{d\ell} Y= \frac{Y(X-Y)(2(5T+1)X +(T+1)^2 Y -4T(T-1))}{T(T+1)^2}.
\end{eqnarray} 
 A priori, the flow equations are valid for arbitrary $(X,Y,T)\in\mathbb{R}^3$. 
 Since $T$ is a ratio of diffusion constants, we require $T>0$. In the present  context stability is required, which translates to $XY > 0$.  

Universality classes are labeled by the fixed points of the flow equations.
Setting the left hand side  of \eqref{4.2} to zero, one arrives at a line of fixed points defined by
\begin{equation}
\label{4.3}
P_\mathrm{f} = (X_\mathrm{f},X_\mathrm{f},T(X_\mathrm{f})), \quad T(X)=\tfrac{1}{2}\big(\sqrt{(1+X^2)^2+12 X^2}-1-X^2\big)
\end{equation}
with $X_\mathrm{f}\in \mathbb{R}\setminus\{0\}$. This result is in complete agreement with our analysis. $P_\mathrm{f}$ has two stable directions and one neutral
direction with eigenvector pointing along the line of fixed points. Starting with general initial data satisfying $X Y> 0$, $T>0$, in the limit $\ell \to \infty$, a unique fixed point is reached. The universality class labeled by $X_\mathrm{f}$ 
is the basin of attraction of $P_\mathrm{f}$ and consists of all flow lines 
ending up at $P_\mathrm{f}$. Thereby the three-dimensional phase space of Eq. \eqref{4.2} is foliated into universality classes, which are two-dimensional surfaces
in three-space. The flow equations in \cite{1993-ertas-kardar} include also terms, not displayed here,
from which one infers that $\chi = \tfrac{1}{2}$ and $z = \tfrac{3}{2}$, in accordance with our findings.

Further items for comparison are the two border lines $X = 0$ and $Y=0$, compare with Section \ref{sec2.8}. For initial $(X,0,T)$
the flow converges to the fixed point $(1,0,1)$. This is in agreement with Section \ref{sec2.8}, where it is argued that $h_1$ will be in the KPZ universality class, while $h_2$ should be Gaussian. However this fixed point has an unstable direction pointing towards $Y<0$. In this sense, the positive $X$-axis is a borderline.
In the unstable regime, $X Y< 0$, there is a single further fixed point given by $(X,Y,T)=(1,-1,1)$, the properties of which are reported in \cite{weinberger2024}. 
The initial data $(0,Y,T)$ flow towards the fixed  point $(0,0,0)$, which indicates diffusive scaling for both components. The analysis in Section \ref{sec2.8} suggests that $h_1$ is Gaussian and $h_2$ has KPZ scaling.  

\begin{figure}[!h]
\centering
    \includegraphics[width=10.0cm]{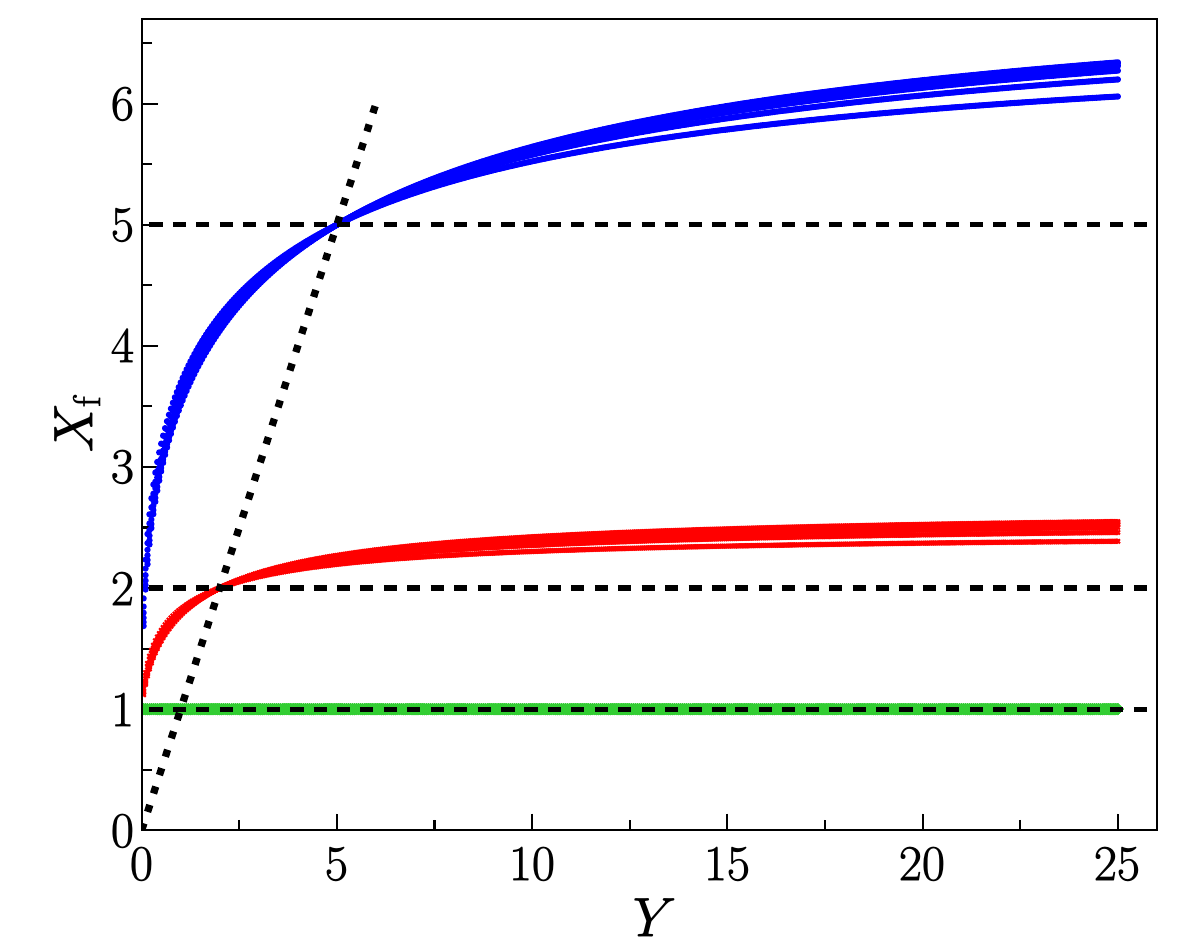}
    \caption{Figure illustrates the RG flowlines (solid lines) for $X=1,2,5$, represented by green, red and blue respectively. The fixed point $X_\mathrm{f}$ is plotted versus $Y$ for a range of $T$ values, i.e., $T=0.4,0.6,0.8,1.0,1.2,1.4$. We notice a relatively small spread while varying $T$. The horizontal lines (dashed) are independent of $T$ and display our predictions. The black dotted line represents $X_f=Y$.}
\label{fig-flow}
\end{figure}

While there is qualitative agreement,
 the surfaces
defining the $X_\mathrm{f}$ universality classes differ from each other. According to our analysis, the surface is the
$Y$-$T$ plane shifted to $X_\mathrm{f}$. For the RG approach one has to compute numerically the basin of attraction of the fixed point with label $X_\mathrm{f}$. 
While this can be done, to have an impression we only plot particular sections.
We consider fixed $X = 1,2,5$, $T= 0.5,1,2$ and solve \eqref{sec2.5} with initial data $P_0 =(X,Y,T)$ allowing for general $Y$. Then 
 $X_\mathrm{f}$ is plotted as a function of $Y$, see Figure \ref{fig-flow}. According to our claim one should have simply the constant function  $X_\mathrm{f} = X$ independent of $Y,T$. The flow equations \eqref{4.2} make a distinct prediction.

In \cite{RDK24} we also studied the two-lane lattice gas of \cite{Schutz}. At half-filling, the model is argued to be in the same universality class as coupled KPZ equations at $(X,X,1)$, where $X$ is related to a coupling coefficient
between the two lanes. Very recently, for the same model, extensive numerical simulations are reported in \cite{arxiv}.
They confirm that the scaling functions vary nontrivially as a function of $X$. Also for short times both studies are in agreement within numerical error bars. However, in \cite{arxiv} much longer time scales are reached and it is concluded, that the scaling exponent of peak 1 is slightly smaller than $\tfrac{3}{2}$ and the one of peak 2 slightly larger than $\tfrac{3}{2}$.


To clarify such discrepancies,
a promising direction is the well-developed 
functional renormalization of KPZ \cite{PhysRevE.84.061128}. Originally its focus was on KPZ in higher dimensions. But the method has been adjusted to one dimension. There are predictions for the Fourier transform of the dynamic correlator. The agreement with the exact solution is impressive. Even the small negative dip in the scaling function is reproduced, see Section VI of  \cite{PhysRevE.84.061128}. Thus an interesting goal for the future is to extend functional RG to coupled KPZ equations in one dimension.

\appendix
\section{Dimensionless units}
\setcounter{equation}{0} 
\label{A}
Let us start from Eq. \eqref{1.3} with $n=2$,
\begin{equation}
    \label{A.01}
    \partial_t h_\alpha =    G^\alpha_{\beta\gamma} (\partial_x h_\beta )(\partial_x h_\gamma) + \tfrac{1}{2} D_{\alpha\beta}\partial_x^2 h_\beta + B_{\alpha\beta}\xi_\beta,
\end{equation}
$\alpha,\beta,\gamma = 1,2$. Imposing the constraint of invariance under interchanging the two components, the coupling matrices acquire the special form
\begin{equation}
    \label{A.02} 
    G^1 = 
    \begin{pmatrix}
        \Gamma_{11} & \Gamma_{12}\\
        \Gamma_{12} & \Gamma_{22}\\
    \end{pmatrix}, \quad
    G^2 = 
    \begin{pmatrix}
        \Gamma_{22} & \Gamma_{12}\\
        \Gamma_{12} & \Gamma_{11}\\
    \end{pmatrix},\quad
    D = 
    \begin{pmatrix}
        d_1 & d_2\\
        d_2 & d_1\\
    \end{pmatrix}, \quad
    B = 
    \begin{pmatrix}
        b_1 & b_2\\
        b_2 & b_1\\
    \end{pmatrix}.
\end{equation}
It is convenient to rotate the equations of motion by $\pi/4$, to say 
\begin{equation}
\label{A.03}
\tilde{h}  = R h, \quad R = 
\frac{1}{\sqrt{2}}\begin{pmatrix}
 1 & -1\\
 1 &1
\end{pmatrix},
\end{equation}
which yields 
\begin{eqnarray}
\label{A.04}
 \hspace{-40pt} &&\partial_t \tilde{h}_1 =    2 \tfrac{1}{\sqrt{2}}(\Gamma_{11}-\Gamma_{22})  (\partial_x \tilde{h}_1)(\partial_x \tilde{h}_2) + \tfrac{1}{2} (d_1 -d_2)\partial_x^2 \tilde{h}_1+ (b_1 -b_2)\xi_1 , \nonumber\\
\hspace{-40pt}	&&\partial_t \tilde{h}_2 = \tfrac{1}{\sqrt{2}}(\Gamma_{11}+\Gamma_{22} - 2\Gamma_{12})  (\partial_x\tilde{h}_1)^2 + 
    \tfrac{1}{\sqrt{2}}(\Gamma_{11}+\Gamma_{22} + 2\Gamma_{12})(\partial_x\tilde{h}_2)^2 \nonumber\\
	&& \hspace{40pt}+  \tfrac{1}{2}(d_1 + d_2)\partial_x^2\tilde{h}_2  +(b_1 +b_2)\xi_2.
    \end{eqnarray}
The computation simplifies by employing that the noise is  invariant under the rotations.

To simplify notation we remove the tilde and abbreviate the coefficients in the obvious way.  Then  $h_\alpha$ satisfies the evolution equations
\begin{eqnarray}
\label{A.1}
    &&\partial_t h_1 =    2 \Gamma_3  (\partial_x h_1)(\partial_x h_2) + \tfrac{1}{2} D_1\partial_x^2 h_1+ \sigma_1\xi_1 , \nonumber\\
	&&\partial_t h_2 =  \Gamma_1  (\partial_xh_1)^2 + \Gamma_2(\partial_xh_2)^2 +  \tfrac{1}{2}D_2\partial_x^2 h_2  +\sigma_2\xi_2.
    \end{eqnarray}  
There are still seven parameters and the goal now is to reduce their number. 
For this purpose we rescale the 
fields by choosing units of time, space, and amplitudes, through setting
\begin{equation}
\label{A.2}
h_\alpha(x,t) = a_\alpha \tilde{h}_\alpha (x/\ell,t/\tau)\quad\mathrm{or}\quad
\tilde{h}_\alpha(x,t) = (a_\alpha)^{-1} h_\alpha (\ell x,\tau t)
\end{equation}
with adjustable coefficients $a_\alpha,\ell,\tau$.
Inserting in \eqref{A.1}, one obtains 
\begin{eqnarray}
\label{A.3}
    &&\partial_t \tilde{h}_1 =    2 \frac{a_2\tau}{\ell^2}\Gamma_3  (\partial_x \tilde{h}_1)(\partial_x \tilde{h}_2) +  \frac{\tau}{2\ell^2}D_1\partial_x^2 \tilde{h}_1+ \frac{\sqrt{\tau}}{a_1\sqrt{\ell}}\sigma_1\xi_1 , \nonumber\\
	&&\partial_t \tilde{h}_2 =  \frac{a_1^2\tau}{a_2\ell^2}\Gamma_1  (\partial_x\tilde{h}_1)^2 + 
	\frac{a_2\tau}{\ell^2}\Gamma_2(\partial_x\tilde{h}_2)^2 +  \frac{\tau}{2\ell^2} D_2\partial_x^2 \tilde{h}_2  +\frac{\sqrt{\tau}}{a_2\sqrt{\ell}}\sigma_2\xi_2.
    \end{eqnarray} 
The parameters $a_1, \ell,\tau$ are chosen such that     
\begin{equation}
\label{A.4}
\frac{\tau}{\ell^2}D_1 = \frac{\tau\sigma_1^2}{a_1^2 \ell}, \quad \frac{\tau}{\ell^2}D_2 =1,\quad  \frac{\sqrt{\tau}}{a_2\sqrt{\ell}}\sigma_2 =1.
\end{equation}      
Then the prefactor of $\tfrac{1}{2}\partial_x^2 \tilde{h}_1$ turns into $(D_1/D_2)$. Inserting this choice in the nonlinear term
yields the successive coefficients of the nonlinearities,
\begin{equation}
\label{A.5}
\frac{a_2 \Gamma_2}{D_2 } \times\Big(\,\frac{\Gamma_3}{\Gamma_2}\,, \,\frac{\sigma_1^2 D_2 \Gamma_1}{\sigma_2^2 D_1\Gamma_2} \,, \,1\,\Big).
\end{equation}  
Setting $a_2 \Gamma_2 = D_2$, the factor in front of the square bracket equals 1. Finally, setting 
\begin{align}
(D_1/D_2) = T, ~ (\Gamma_3/\Gamma_2) = X,~ 
(\sigma_1^2 D_2\Gamma_1/\sigma_2^2 D_1\Gamma_2) = Y 
\end{align}
results in Eq. \eqref{1.4}, as claimed.

In Sections \ref{sec2.6} and \ref{sec3.1} the following identity is used. We start from
\begin{equation}\label{A.6}
    \begin{aligned}
	\partial_t h_1 &=    2\mathsfit{b} X (\partial_x h_1)(\partial_x h_2) + \tfrac{1}{2} \mathsfit{d}T \partial_x^2 h_1+ \sqrt{\mathsfit{d}T}\xi_1 , \\
	\partial_t h_2 &= \mathsfit{b} Y  (\partial_x h_1)^2 + \mathsfit{b}(\partial_x h_2)^2 +  \tfrac{1}{2}\mathsfit{d}\partial_x^2 h_2  + \sqrt{\mathsfit{d}}\xi_2.
    \end{aligned}
\end{equation}
Setting $a_1= \mathsfit{d}/\mathsfit{b} $, $a_2=\mathsfit{d}/\mathsfit{b}$, $\ell= \mathsfit{d}^2/\mathsfit{b}^2$, $\tau= \mathsfit{d}^3/\mathsfit{b}^4$, the transformed fields 
satisfy\begin{equation}\label{A.7}
    \begin{aligned}
	\partial_t h_1 &=    2X (\partial_x h_1)(\partial_x h_2) + \tfrac{1}{2} T \partial_x^2 h_1+ \sqrt{T}\xi_1 , \\
	\partial_t h_2 &=  Y  (\partial_x h_1)^2 + (\partial_x h_2)^2 +  \tfrac{1}{2}\partial_x^2 h_2  + \xi_2.
    \end{aligned}\bigskip
\end{equation}
\textbf{Acknowledgments}. 
HS thanks Jacopo DeNardis, Ziga Krajnik, Joachim Krug,  and Harvey Weinberger for most instructive discussions.  AD, MK and HS thank the VAJRA faculty scheme (No. VJR/2019/000079) from the Science and Engineering Research Board (SERB), Department of Science and Technology, Government of India.  AD and MK acknowledges the Department of Atomic Energy, Government of India, for their support under Project No. RTI4001. AD acknowledges the J.C. Bose Fellowship (JCB/2022/000014) of the Science and Engineering Research Board of the Department of Science and Technology, Government of India.


\section*{References}
\bibliographystyle{unsrt}
\bibliography{2kpz-refs-new.bib}

\begin{thebibliography}{10}

\bibitem{1986-kardar--zhang}
Mehran Kardar, Giorgio Parisi, and Yi-Cheng Zhang.
\newblock Dynamic scaling of growing interfaces.
\newblock {\em Physical Review Letters}, 56:889, 1986.

\bibitem{forster}
Dieter Forster, David~R. Nelson, and Michael~J. Stephen.
\newblock Large-distance and long-time properties of a randomly stirred fluid.
\newblock {\em Phys. Rev. A}, 16:732, 1977.

\bibitem{Krug1977}
Joachim Krug.
\newblock Scaling relation for a growing interface.
\newblock {\em Phys. Rev. A}, 36:5465, 1987.

\bibitem{medina}
Ernesto Medina, Terence Hwa, Mehran Kardar, and Yi-Cheng Zhang.
\newblock Burgers equation with correlated noise: Renormalization-group
  analysis and applications to directed polymers and interface growth.
\newblock {\em Phys. Rev. A}, 39:3053, 1989.

\bibitem{2015-quastel-spohn}
Jeremy Quastel and Herbert Spohn.
\newblock The one-dimensional {KPZ} equation and its universality class.
\newblock {\em Journal of Statistical Physics}, 160:965, 2015.

\bibitem{ps2002}
Michael Pr{\"a}hofer and Herbert Spohn.
\newblock Scale invariance of the {PNG} droplet and the {A}iry process.
\newblock {\em Journal of Statistical Physics}, 108:1071, 2002.

\bibitem{2018-takeuchi}
Kazumasa~A. Takeuchi.
\newblock {An appetizer to modern developments on the {K}ardar-{P}arisi-{Z}hang
  universality class}.
\newblock {\em Physica A: Statistical Mechanics and its Applications}, 504:77,
  2018.
\newblock {Lecture Notes of the 14th International Summer School on Fundamental
  Problems in Statistical Physics}.

\bibitem{weinberger2024}
Harvey Weinberger, Paolo Comaron, and Marzena~H. Szymańska.
\newblock Multicomponent {K}ardar-{P}arisi-{Z}hang universality in degenerate
  coupled condensates.
\newblock {\em arXiv:2411.07095}, 2024.

\bibitem{1993-ertas-kardar}
Deniz Erta\ifmmode~\mbox{\c{s}}\else \c{s}\fi{} and Mehran Kardar.
\newblock Dynamic relaxation of drifting polymers: A phenomenological approach.
\newblock {\em Physical Review E}, 48:1228, 1993.

\bibitem{2017-chakraborty--barma}
Shauri Chakraborty, Sakuntala Chatterjee, and Mustansir Barma.
\newblock Ordered phases in coupled nonequilibrium systems: Static properties.
\newblock {\em Physical Review E}, 96:022127, 2017.

\bibitem{2019-chakraborty--barma}
Shauri Chakraborty, Sakuntala Chatterjee, and Mustansir Barma.
\newblock Dynamics of coupled modes for sliding particles on a fluctuating
  landscape.
\newblock {\em Physical Review E}, 100:042117, 2019.

\bibitem{2023-nardis--vasseur}
Jacopo De~Nardis, Sarang Gopalakrishnan, and Romain Vasseur.
\newblock Nonlinear fluctuating hydrodynamics for {Kardar-Parisi-Zhang} scaling
  in isotropic spin chains.
\newblock {\em Physical Review Letters}, 131:197102, 2023.

\bibitem{2021-scheie--tennant}
A.~Scheie, N.~E. Sherman, M.~Dupont, S.~E. Nagler, M.~B. Stone, G.~E. Granroth,
  J.~E. Moore, and D.~A. Tennant.
\newblock Detection of {Kardar–Parisi–Zhang} hydrodynamics in a quantum
  {Heisenberg} spin-1/2 chain.
\newblock {\em Nature Physics}, 17:726, 2021.

\bibitem{SR1}
Rangan Lahiri and Sriram Ramaswamy.
\newblock Are steadily moving crystals unstable?
\newblock {\em Phys. Rev. Lett.}, 79:1150, 1997.

\bibitem{SR2}
Rangan Lahiri, Mustansir Barma, and Sriram Ramaswamy.
\newblock Strong phase separation in a model of sedimenting lattices.
\newblock {\em Phys. Rev. E}, 61:1648, 2000.

\bibitem{2011-grisi-schutz}
Rafael~M Grisi and Gunter~M Sch{\"u}tz.
\newblock Current symmetries for particle systems with several conservation
  laws.
\newblock {\em Journal of Statistical Physics}, 145:1499, 2011.

\bibitem{2013-ferrari--spohn}
Patrik~L Ferrari, Tomohiro Sasamoto, and Herbert Spohn.
\newblock Coupled {Kardar-Parisi-Zhang} equations in one dimension.
\newblock {\em Journal of Statistical Physics}, 153:377, 2013.

\bibitem{2013-mendl-spohn}
Christian~B. Mendl and Herbert Spohn.
\newblock Dynamic correlators of {Fermi-Pasta-Ulam} chains and nonlinear
  fluctuating hydrodynamics.
\newblock {\em Physical Review Letters}, 111:230601, 2013.

\bibitem{2014-das--spohn}
Suman~G. Das, Abhishek Dhar, Keiji Saito, Christian~B. Mendl, and Herbert
  Spohn.
\newblock Numerical test of hydrodynamic fluctuation theory in the
  {Fermi-Pasta-Ulam} chain.
\newblock {\em Physical Review E}, 90:012124, 2014.

\bibitem{Dolai2024}
Pritha Dolai, Aditi Simha, and Abhik Basu.
\newblock {Kardar-Parisi-Zhang universality in two-component driven diffusive
  models: symmetry and renormalization group perspectives}.
\newblock {\em Phys. Rev. E}, 109:064122, 2024.

\bibitem{2014-spohn}
Herbert Spohn.
\newblock Nonlinear fluctuating hydrodynamics for anharmonic chains.
\newblock {\em Journal of Statistical Physics}, 154:1191, 2014.

\bibitem{RDK24}
Dipankar Roy, Abhishek Dhar, Konstantin Khanin, Manas Kulkarni, and Herbert
  Spohn.
\newblock Universality in coupled stochastic {B}urgers systems with degenerate
  flux {J}acobian.
\newblock {\em Journal of Statistical Mechanics: Theory and Experiment},
  2024:033209, 2024.

\bibitem{Krug01041997}
Joachim Krug.
\newblock Origins of scale invariance in growth processes.
\newblock {\em Advances in Physics}, 46:139, 1997.

\bibitem{JanssenPRL}
H.~K. Janssen.
\newblock On critical exponents and the renormalization of the coupling
  constant in growth models with surface diffusion.
\newblock {\em Phys. Rev. Lett.}, 78:1082, 1997.

\bibitem{FH17}
Tadahisa Funaki and Masato Hoshino.
\newblock A coupled {KPZ} equation, its two types of approximations and
  existence of global solutions.
\newblock {\em Journal of Functional Analysis}, 273:1165, 2017.

\bibitem{FH15}
Tadahisa Funaki.
\newblock Infinitesimal invariance for the coupled {KPZ} equations.
\newblock In Catherine Donati-Martin, Antoine Lejay, and Alain Rouault,
  editors, {\em In Memoriam Marc Yor - S{\'e}minaire de Probabilit{\'e}s
  XLVII}, page~37. Springer International Publishing, Cham, 2015.

\bibitem{H23}
Kohei Hayashi.
\newblock Derivation of coupled {KPZ} equations from interacting diffusions
  driven by a single-site potential.
\newblock {\em arXiv:2208.05374}, 2023.

\bibitem{schmidt}
Jan de~Gier, Andreas Schadschneider, Johannes Schmidt, and Gunter~M. Sch\"utz.
\newblock Kardar-{P}arisi-{Z}hang universality of the {N}agel-{S}chreckenberg
  model.
\newblock {\em Phys. Rev. E}, 100:052111, 2019.

\bibitem{Schutz}
Vladislav Popkov and Gunter~M. Sch\"utz.
\newblock Unusual shock wave in two-species driven systems with an umbilic
  point.
\newblock {\em Phys. Rev. E}, 86:031139, 2012.

\bibitem{arxiv}
Johannes Schmidt, Žiga Krajnik, and Vladislav Popkov.
\newblock Universality in many-body driven systems with an umbilic point.
\newblock {\em arXiv:2504.04304}, 2025.

\bibitem{PhysRevE.84.061128}
L\'eonie Canet, Hugues Chat\'e, Bertrand Delamotte, and Nicol\'as Wschebor.
\newblock Nonperturbative renormalization group for the
  {K}ardar-{P}arisi-{Z}hang equation: General framework and first applications.
\newblock {\em Phys. Rev. E}, 84:061128, 2011.

\end{thebibliography}

\end{document}